\documentclass[pra,aps,superscriptaddress,twocolumn,color]{revtex4-2}

\usepackage{amsmath}
\usepackage{amsfonts}
\usepackage{bm}
\usepackage{graphicx}
\usepackage{array,color}
\usepackage{ulem}

\begin{document}
\title{Neural-network quantum states for solving few-body problems:
  application to Efimov physics}

\author{Sora Yokoi}
\affiliation{Department of Engineering Science, University of
Electro-Communications, Tokyo 182-8585, Japan}

\author{Shimpei Endo}
\affiliation{Department of Engineering Science, University of
Electro-Communications, Tokyo 182-8585, Japan}

\author{Hiroki Saito}
\affiliation {Department of Engineering Science, University of
Electro-Communications, Tokyo 182-8585, Japan}

\date{\today}

\begin{abstract}
Neural-network quantum states have recently emerged as a powerful
method for solving quantum many-body problems, with notable successes
in lattice systems.
Here, we extend this approach to strongly interacting few-body
problems in continuous space, and demonstrate its capability by
computing the Efimov states and associated few-body bound states.
Using a fully connected feedforward neural network with Jacobi
coordinates as inputs, combined with a projection method, we compute
the ground and first excited states for three- to six-body systems of
identical bosons at unitarity, as well as a mass-imbalanced fermionic
system consisting of two identical fermions and a third particle.
The obtained energies of the ground and first excited states
agree well with previously reported results.
Furthermore, the proposed approach also reproduces key features of
Efimov states, including the discrete scale invariance, the
characteristic geometric structure of the wave function, and the
critical-mass behavior in mass-imbalanced fermionic systems.
Our method can be readily applied to a broad class of strongly
correlated few-body problems in continuous space.
\end{abstract}

\maketitle

\section{Introduction}
\label{s:intro}

Numerical calculations for quantum many-body systems face fundamental
limitations due to the exponential growth of the Hilbert space
with an increase in the degrees of freedom.
To overcome this problem, diverse numerical approaches have been
developed, including quantum Monte Carlo methods~\cite{QMC},
density-matrix renormalization groups~\cite{DMRG}, and tensor
networks~\cite{TN}.
In addition to these approaches, novel methods based on artificial
neural networks have recently attracted much interest~\cite{NQS,
  review1, review2, review3, review4}.
Leveraging the flexibility and adaptability of neural networks for
representing various features, many-body wave functions are
encoded on neural networks with a comparatively small number of
network parameters, which are optimized using machine learning
techniques.
The method of neural-network quantum states was first applied to
spatially discrete problems, such as spins and particles on
lattices~\cite{NQS, BHM, Nomura1, Cai, BHM2}, and then extended
to continuous-space problems, such as
interacting bosons~\cite{boson1, boson2, boson3, boson4, boson5},
fermionic systems~\cite{fermi1, fermi2, fermi3, fermi4, fermi5,
  fermi6, fermi7, fermi8},
and nucleons in nuclei~\cite{nuclei1, nuclei2, nuclei3, nuclei4,
  nuclei5, nuclei6, nuclei7}.

Neural-network quantum states are not only powerful tools for ground
states, but also useful for exploring excited states~\cite{Choo, Nomura,
  Viteritti, Entwistle, Pfau, Li}.
An excited state belonging to a symmetry sector different from the
ground state can be simply obtained by minimizing the energy by fixing
quantum numbers, such as momentum or magnetization.
Even if the desired excited state cannot be distinguished from the
ground state by quantum numbers, one can still compute the excited
state by projecting it to the Hilbert space orthogonal to the ground
state.
Using these procedures, low-lying excitations of the Heisenberg model,
the Bose-Hubbard model~\cite{Choo}, and the $J_1$-$J_2$
model~\cite{Nomura, Viteritti} have been calculated.
More recently, multiple excited states of benzene-scale molecules have
been computed by regarding them as the ground state of an extended
system~\cite{Pfau}.

While neural-network quantum states have been successfully applied to
a variety of quantum systems, it remains unclear to what extent they
can serve as viable tools for strongly correlated quantum few-body
problems, which are of pivotal importance in nuclear physics and cold
atomic systems.
In nuclear physics, a rich variety of nuclear structures and reactions
emerge from strong interactions between protons and
neutrons~\cite{ring1980nuclear}.
In cold atoms, strongly correlated few- and many-body systems can be
realized by tuning the $s$-wave scattering length between the atoms
via a Feshbach
resonance~\cite{inouye1998observation,chin2010feshbach}, enabling the
experimental observation of the Efimov
states~\cite{efimov, efimov1973energy, kraemer2006evidence,
  Braaten2006259, RevModPhys.89.035006, naidon, d2018few,
  endoepelbaumQcl2024}.
The Efimov states are peculiar three-body bound states characterized
by Borromean binding and discrete scale invariance, that manifest as a
quantum anomaly in the underlying field
theory~\cite{PhysRevLett.82.463, Braaten2006259, RevModPhys.89.035006,
  naidon, d2018few, endoepelbaumQcl2024}.
To understand the physical nature of the Efimov states and to gain
insight into many-body physics from a few-body perspective, few-body
systems with large $s$-wave scattering lengths have been extensively
studied both
experimentally~\cite{kraemer2006evidence, Zaccanti2009,
  PhysRevLett.103.163202, doi:10.1126/science.1182840,
  PhysRevLett.102.140401, Zenesini_2013, PhysRevLett.112.250404,
  PhysRevLett.113.240402} and
theoretically~\cite{PhysRevLett.83.1751, von2009signatures,
  vonStecher2010, finite,PhysRevA.86.012502, PhysRevA.90.052514}.
These studies have revealed that, in addition to three-body bound
states, four-, five-, and six-body states tied to Efimov trimers also
appear universally~\cite{doi:10.1126/science.1182840,
  PhysRevLett.102.140401, Zenesini_2013, von2009signatures,
  vonStecher2010, finite,PhysRevA.90.052514}.
This hierarchy of few-body clusters provides an important step toward
a unified understanding of how few-body systems evolve into many-body
systems as the particle number increases~\cite{science.279.5359.2083,
  Wenz_2013, vonStecher2010, finite,PhysRevA.92.033626,
  Levinsen_2017}.

As Efimov binding is a highly quantum-mechanical phenomenon that
defies classical description, it is an ideal testbed for benchmarking
novel numerical methods.
Indeed, strongly correlated few-body problems have been accurately
solved with high precision using the correlated Gaussian
basis~\cite{Blume, Daily,von2009signatures}, adiabatic hyper-spherical
expansion~\cite{von2009signatures}, Gaussian
expansion~\cite{PhysRevA.86.012502,PhysRevA.90.052514}, and
hyper-spherical harmonics expansion~\cite{finite}.
As the Efimov states appear not only in cold atoms but also
universally in a wide range of systems with large $s$-wave scattering
lengths, such as weakly bound nuclear systems~\cite{AnnRev_HamPlatt,
  endoepelbaumQcl2024, PhysRevLett.100.192502,
  PhysRevLett.120.052502},
$^4$He clusters~\cite{Kunitski}, and magnetic
systems~\cite{nishida2013efimov}, establishing a neutral-network
method for Efimov physics would lay the foundation toward a universal
computational framework for exploring strongly correlated few- and
many-body problems across various physical fields and scales.

Here, we develop a neural-network method for computing not only the
ground state but also excited states in strongly interacting few-body
systems in uniform space and demonstrate that it accurately captures
the Efimov states and their associated few-body clusters.
Our implementation employs a fully connected feedforward neural
network~\cite{BHM,Cai} with suitably chosen Jacobi coordinates as
inputs.
The ground state is obtained by minimizing the variational energy with
respect to the network parameters, and the first excited state is
computed using the orthogonal projection method~\cite{Choo}.
For three- to six-body systems of identical bosons at the unitary
limit, as well as for mass-imbalanced three-body fermionic systems
composed of two identical fermions and another particle, the ground
and first excited states are robustly obtained.
The results are in excellent agreement with previous
results~\cite{finite,Blume,Daily}, achieving comparable or 
better accuracy.
For three-body systems, we reproduce the expected universal
features of the Efimov states, such as the discrete scale invariance,
characteristic wave-function structures, and critical behavior as the
mass ratio approaches $13.606\ldots$ for
fermions~\cite{efimov1973energy,PhysRevA.67.010703}.
Our work thus establishes the neural-network method as a highly
accurate and versatile method for exploring strongly interacting
few-body problems.

The rest of this paper is organized as follows.
Section~\ref{s:NQS} explains the method of neural-network quantum
states.
Sections~\ref{s:bose} and \ref{s:fermi} study systems of identical
bosons and two identical fermions with one particle, respectively.
In each of these sections, the method and numerical results are
presented.
Section~\ref{s:conc} gives the conclusions and suggestions for future
research.


\section{Neural-Network Quantum States}
\label{s:NQS}

To represent a wave function, we use a fully-connected feedforward
neural network, which consists of an input layer, hidden layers, and
an output layer~\cite{Goodfellow}.
The input layer receives a real-valued vector,
\begin{equation} \label{uin}
\bm{u}^{\mathrm{in}} = \left(
u^{\mathrm{in}}_1, u^{\mathrm{in}}_2, \dots,
u^{\mathrm{in}}_{N_{\mathrm{in}}}
\right),
\end{equation}
constructed from the particle coordinates, where $N_{\mathrm{in}}$
denotes the number of input values.
We will specify later how to construct $\bm{u}^{\mathrm{in}}$ from the
particle coordinates in a uniform space.
The neural network contains $L_h$ hidden layers.
The units in the $\ell$-th hidden layer ($\ell = 1, 2, \dots, L_h$)
are defined recursively as
\begin{equation}
u^{(\ell)}_i =
\sum_{j=1}^{N_{\ell-1}}
W^{(\ell)}_{ji}\,
f\left( u^{(\ell-1)}_j \right)
+ b^{(\ell)}_i,
\end{equation}
where $N_{\ell}$ denotes the number of units in the $\ell$-th layer,
$W^{(\ell)}$ is a real $N_{\ell-1} \times N_\ell$ matrix,
and $b^{(\ell)}$ is a real $N_\ell$ component vector.
The input layer corresponds to $\ell = 0$.
As an activation function $f(x)$, we adopt the SiLU (Sigmoid Linear
Unit) function:
\begin{equation}
f(x) = \frac{x}{1 + e^{-x}}.
\end{equation}
We found that this activation function allows more stable and accurate
calculations than those obtained using other activation functions,
such as the ReLU (Rectified Linear Unit) and sigmoid functions.
The output layer gives the following two real values:
\begin{equation}
u^{\mathrm{out}}_k =
\sum_{i=1}^{N_{L_h}} W^{(L_h)}_{ik}
f\left(u^{(L_h)}_i\right)
\qquad (k=1,2).
\end{equation}
In this paper, we use a neural network with $L_h = 3$ and $N_\ell =
32$.
Finally, the network outputs a single value $A$:
\begin{equation} \label{A}
A = u^{\mathrm{out}}_1
\exp\left(
u^{\mathrm{out}}_2
\right),
\end{equation}
so that $A$ can take either sign with exponential nonlinearity.
The input values $\bm{u}_{\rm in}$ are made from the particle
coordinates $X$.
The many-body trial wave function $\Psi$ is constructed using the
network output $A(X, W)$ in combination with $X$, where $W$ denotes
the network parameters.
The wave function $\Psi$ therefore depends on both $X$ and $W$.

According to the variational principle, the expectation value $E(W)$
of the Hamiltonian $\hat{H}$ for the trial wave function $\Psi$
is not less than the true ground-state energy $E_0$:
\begin{equation}
E(W) =
\frac{\int \Psi^* \hat{H} \Psi dX}{\int |\Psi|^2 dX}
\ge E_0.
\end{equation}
We minimize $E(W)$ with respect to $W$ using the variational Monte
Carlo method.
In the following sections, we rewrite $E(W)$ in a form generally
given by
\begin{equation} \label{E2}
E(W) = \frac{\int F(P, W) dP}{\int |\Psi(P, W)|^2 dP},
\end{equation}
where $F$ is a function constructed from the network output $A$ and
the Jacobi coordinates $P$ of the particle positions.
By taking the Metropolis-Hastings sampling of the coordinates $P$ with
a probability distribution
\begin{equation}
p(P, W) = \frac{|\Psi(P, W)|^2}{\int |\Psi(P, W)|^2 dP},
\end{equation}
we can evaluate the multidimensional integration in Eq.~(\ref{E2}) as
\begin{eqnarray}
  E(W) & = & \int p(P, W) \frac{F(P, W)}{|\Psi(P, W)|^2} dP
  \nonumber \\
  & \simeq & \frac{1}{N_s} \sum_{n=1}^{N_s}
  \frac{F(P_n, W)}{|\Psi(P_n, W)|^2}, 
\end{eqnarray}
where $N_s$ is the number of samples.
Similarly, the gradient of the energy with respect to the network
parameters is obtained as
\begin{equation} \label{grad}
  \frac{\partial E}{\partial W} \simeq
  \frac{1}{N_s} \sum_{n=1}^{N_s} \frac{1}{|\Psi|^2} \left(
  \frac{\partial F}{\partial W}
  - E \frac{\partial |\Psi|^2}{\partial W} \right).
\end{equation}
Using this gradient, we update the network parameters using the Adam
optimizer~\cite{Adam} to minimize the energy.
The learning rate in the Adam optimizer is typically chosen to be
$10^{-3}$--$10^{-4}$.

We use the projection method to obtain the excited state.
Namely, we define the following wave function~\cite{Choo}:
\begin{equation} \label{psiex}
\Psi(P, W) = \Psi_1(P, W) - \lambda(W) \Psi_0(P, W_0),
\end{equation}
where $\Psi_0(P, W_0)$ is the ground-state wave function obtained
using the above method and $\Psi_1(P, W)$ is a wave function
constructed from a network that is different from that used for the
ground state.
The parameter $\lambda$ in Eq.~(\ref{psiex}) is given by
\begin{equation}
\lambda(W) = \frac{
\int \Psi_0^*(P, W_0) \Psi_1(P, W) dP
}{
\int |\Psi_0(P, W_0)|^2 dP
},
\end{equation}
which is also evaluated using Monte Carlo sampling.
The wave function $\Psi$ in Eq.~(\ref{psiex}) is orthogonal to the
ground state $\Psi_0$.
By minimizing the energy with respect to the network parameters $W$, we
obtain the first excited state.
The gradient of the energy can be numerically calculated using the
automatic differentiation provided by a machine-learning framework.


\section{\label{s:bose}Bosonic Systems}

\subsection{\label{s:bose_method}Method}

We first apply our neural-network method to systems of identical $N$
bosons with mass $m$ in three-dimensional free space.
The position of the $i$-th particle is denoted by the Cartesian
coordinate $\bm{r}_i \in \mathbb{R}^3$.
The $i$-th and $j$-th particles interact with each other through the
two-body potential $V(r_{ij})$, where $r_{ij} = |\bm{r}_i-\bm{r}_j|$.
The Hamiltonian for the system is given by
\begin{equation} \label{H}
\hat{H} =
-\frac{\hbar^2}{2m}\sum_{i=1}^{N} \frac{\partial^2}{\partial \bm{r}_i^2}
+ \sum_{i<j}
V \left( r_{ij} \right).
\end{equation}
The system has translational symmetry and we eliminate the
center-of-mass motion using the Jacobi coordinates,
\begin{equation}
\begin{aligned}
\bm{\rho}_{N-j} ={}&
\sqrt{\frac{2j}{j+1}}
\Biggl[
\bm{r}_{j+1} - \frac{1}{j}
\left(
\bm{r}_1 + \bm{r}_2 + \cdots + \bm{r}_j
\right)
\Biggr],
\\
&\qquad
j=1,2,\dots,N-1,
\end{aligned}
\end{equation}
and the center-of-mass coordinate,
\begin{equation}
\bm{\rho}_{\mathrm{cm}} =
\sqrt{\frac{2}{N(N+1)}}
\left(\bm{r}_1+\bm{r}_2+\cdots+\bm{r}_N\right).
\end{equation}
Transforming from the Cartesian coordinates to
$\{\bm{\rho}_{\mathrm{cm}}, \bm{\rho}_1, \bm{\rho}_2, \dots,
\bm{\rho}_{N-1}\}$,
we can decompose the Hamiltonian $\hat{H}$ as
\begin{equation}
\hat{H} = -\frac{\hbar^2}{m} \frac{1}{N+1} \frac{\partial^2}{\partial
  \bm{\rho}_{\mathrm{cm}}^2} + \hat{H}',
\end{equation}
where
\begin{equation} \label{Hprime}
\hat{H}' = -\frac{\hbar^2}{m} \sum_{j=1}^{N-1}
\frac{\partial^2}{\partial \bm{\rho}_j^2}
+ \sum_{i<j} V \left( r_{ij} \right).
\end{equation}
The relative distance $r_{ij}$ in the potential terms can only be
expressed by $\bm{\rho}_1, \bm{\rho}_2, \dots$, and $\bm{\rho}_{N-1}$,
and therefore, the decomposed part $\hat{H}'$ does not involve
$\bm{\rho}_{\mathrm{cm}}$.
We therefore only consider $\hat{H}'$ and minimize its expectation
value in the following.

The Hamiltonian $\hat{H}'$ has rotational symmetry.
We only consider the rotationally symmetric wave functions, which
should be functions of the magnitudes of the Jacobi vectors,
\begin{equation}
\rho_i = |\boldsymbol{\rho}_i|
\qquad (i = 1,2,\dots,N-1),
\end{equation}
and the angles between pairs of the Jacobi vectors,
\begin{equation}
\theta_{ij} = \arccos
\left(
\frac{
\boldsymbol{\rho}_i \cdot \boldsymbol{\rho}_j
}{
|\boldsymbol{\rho}_i|\,|\boldsymbol{\rho}_j|
}
\right)
\qquad (i<j).
\end{equation}
We therefore regard these continuous variables as the neural network's
input vector in Eq.~(\ref{uin}) as
\begin{equation}
\boldsymbol{u}^{\mathrm{in}} =
\left(
\rho_1,\rho_2,\dots,\rho_{N-1},
\theta_{12},\theta_{13},\dots
\right).
\end{equation}
With this choice of inputs, the translational and rotational degrees
of freedom are removed at the input level.
The neural network can thus efficiently optimize its parameters while
keeping the symmetries of the quantum states.
In general, a reduction of the number of inputs improves the stability
and efficiency of optimization.
On the other hand, the particle-exchange symmetry required for
identical bosons is not explicitly imposed.
Nevertheless, we numerically find that the obtained wave functions are
sufficiently accurate, almost satisfying the particle-exchange
symmetry~\cite{boson1}.

To accelerate the convergence, we explicitly incorporate the two-body
correlations into the trial wave function.
Let $\phi(r)$ be a function expressing the short-range behavior of
two particles.
We construct the trial wave function $\Psi$ using $\phi(r)$ as
\begin{equation} \label{phiA}
\Psi = \varphi A,
\end{equation}
where 
\begin{equation} \label{varphi}
\varphi = \prod_{i<j} \phi(r_{ij})
\end{equation}
is the product of the two-body correlation $\phi(r)$ for all particle
pairs and $A$ is the network output in Eq.~(\ref{A}).
The multiplication of the two-body correlation $\phi$ assists the
neural network in expressing the short-range behavior of the wave
function induced by two-body potentials, by which abrupt variations of
the network output can be mitigated.
Using the wave function in Eq.~(\ref{phiA}), the energy integral of
the Hamiltonian in Eq.~(\ref{H}) becomes
\begin{eqnarray} \label{cal}
  & & \int dX \left[ -\frac{\hbar^2}{2m} \sum_{i=1}^N \Psi
  \frac{\partial^2}{\partial \bm{r}_i^2} \Psi
  + \sum_{i<j} V(r_{ij}) \Psi^2 \right]
\nonumber \\
& = & \int dX \left[ -\frac{\hbar^2}{2m} \sum_{i=1}^N \varphi A
  \left( 2 \frac{\partial\varphi}{\partial\bm{r}_i} \cdot
  \frac{\partial A}{\partial\bm{r}_i}
  + \varphi \frac{\partial^2 A}{\partial \bm{r}_i^2} \right)
  + V_{\rm eff} \Psi^2 \right]
\nonumber \\
& = & \int dX \left[ \frac{\hbar^2}{2m} \sum_{i=1}^N \varphi^2
  \left( \frac{\partial A}{\partial\bm{r}_i} \right)^2
  + V_{\rm eff} \Psi^2 \right],
\end{eqnarray}
where $dX = d\bm{r}_1 \cdots d\bm{r}_N$.
The effective potential $V_{\rm eff}$ on the second line of
Eq.~(\ref{cal}) is defined as
\begin{eqnarray}
  -\frac{\hbar^2}{2m} \sum_{i=1}^N
  \frac{\partial^2\varphi}{\partial \bm{r}_i^2}
  & = & -\frac{\hbar^2}{2m} \sum_{i=1}^N \sum_{j \neq i} \Biggl[
    \frac{1}{\phi(r_{ij})}
    \frac{\partial^2 \phi(r_{ij})}{\partial \bm{r}_i^2}
\nonumber \\
    & & + \sum_{k \neq i, j} \bm{b}_{ij} \cdot \bm{b}_{ik}
    \Biggr] \varphi
  \nonumber \\
  & \equiv & -\sum_{i<j} V(r_{ij}) \varphi + V_{\rm eff} \varphi,
  \label{cal2}
\end{eqnarray}
where
\begin{equation} \label{b}
  \bm{b}_{ij} = \frac{1}{\phi(r_{ij})}
  \frac{\partial\phi(r_{ij})}{\partial \bm{r}_i}.
\end{equation}
Eliminating the center-of-mass coordinate using the Jacobi
coordinates, we can rewrite Eq.~(\ref{cal}) as
\begin{equation} \label{finalH}
\int \Psi \hat{H}' \Psi dP =
\int dP \left[ \frac{\hbar^2 \varphi^2}{m} \sum_{j=1}^{N-1}
  \left( \frac{\partial A}{\partial\bm{\rho}_j} \right)^2
  + V_{\rm eff} \Psi^2 \right],
\end{equation}
where $dP = d\bm{\rho}_1 d\bm{\rho}_2 \cdots d\bm{\rho}_{N-1}$.

The expectation value of the energy $\int \Psi \hat{H}' \Psi dP / \int
\Psi^2 dP$ is minimized using the method described in Sec.~\ref{s:NQS},
where the function $F$ in Eq.~(\ref{E2}) corresponds to
\begin{equation}
F(P, W) = \frac{\hbar^2 \varphi^2(P)}{m} \sum_{j=1}^{N-1}
  \left[ \frac{\partial A(P, W)}{\partial\bm{\rho}_j} \right]^2
  + V_{\rm eff}(P) \Psi^2(P, W).
\end{equation}
The sampling of the particle positions is performed using the
Metropolis-Hastings algorithm.
Initially, the particle positions $P$ in the Jacobi coordinates are set
to be random.
In each sampling, a random displacement vector $\delta P$ is
generated, which obeys the normal distribution with standard
deviation $\sigma$.
The value of $\sigma / b$ is typically taken to be 1 for the ground
states and 2--10 for the first excited states, where $b$ is a typical
length scale of the two-body interaction.
If $|\Psi(P + \delta P, W)|^2 \geq |\Psi(P, W)|^2$, the displacement
of the particle position is accepted; otherwise, it is accepted
with the probability $|\Psi(P + \delta P, W)|^2 / |\Psi(P, W)|^2$.
In the early stage of the optimization, samples sometimes diverge to
infinity during the random walk.
To suppress such unphysical divergence of particle configurations, we
add an auxiliary harmonic potential~\cite{boson1},
\begin{equation} \label{aux}
\frac{m \omega^2}{4} \sum_{i=1}^{N-1} \rho_i^{2},
\end{equation}
to the Hamiltonian with $\omega \sim 10^{-1}$-$10^{-3} \hbar /
(mb^2)$.
After the first 1000 updates, the auxiliary harmonic potential is
removed to obtain the results in free space.

In the numerical calculations shown in the next subsection, we employ
the P\"oschl-Teller potential for the two-body interaction potential,
\begin{equation}
  V(r) = -\frac{2\hbar^2}{m b^2} \frac{1}{\cosh^2 \left(\frac{r}{b}\right)},
\end{equation}
where $b$ characterizes the range of the interaction.
The potential strength $2\hbar^2/mb^2$ is chosen such that the first
$s$-wave two-body bound state is about to appear, corresponding to an
infinite $s$-wave scattering length (i.e., the unitary limit), which
provides a suitable condition for exploring Efimov physics.
For this potential, the two-body Schr\"odinger equation can be
solved analytically, and the zero-energy solution is obtained
as~\cite{PoschlTeller1933,flugge2012practical}
\begin{equation} \label{sol}
\phi(r) = \frac{1}{r} \tanh{\frac{r}{b}}.
\end{equation}
We employ this solution for the two-body correlation $\phi$ in
Eq.~(\ref{varphi}).
The second derivative
in Eq.~(\ref{cal2}) then becomes
\begin{equation}
-\frac{\hbar^2}{2m} \sum_{i=1}^N \sum_{j \neq i}
    \frac{\partial^2 \phi(r_{ij})}{\partial \bm{r}_i^2} =
    -\sum_{i<j} V(r_{ij}) \phi(r_{ij}),
\end{equation}
and exactly cancels the potential term of $V(r_{ij})$ on the final
line, which yields the simple form of the effective potential as
\begin{equation} \label{Veff}
  V_{\rm eff} = -\frac{\hbar^2}{2m} \sum_{i=1}^N \sum_{j \neq i}
  \sum_{k \neq i, j} \bm{b}_{ij} \cdot \bm{b}_{ik}
\end{equation}
with Eq.~(\ref{b}) being
\begin{equation} \label{b2}
  \bm{b}_{ij} =
  \left( -\frac{1}{r_{ij}} + \frac{2}{b \sinh\frac{2r_{ij}}{b}} \right)
  \frac{\bm{r}_i - \bm{r}_j}{r_{ij}}.
\end{equation}
Thus, the two-body potential is eliminated from the Hamiltonian in
Eq.~(\ref{cal}) and the effective potential appears instead.
For example, in the case of $N = 3$, $V_{\rm eff} =
-\hbar^2 (\bm{b}_{12} \cdot \bm{b}_{13} + \bm{b}_{21} \cdot \bm{b}_{23}
+ \bm{b}_{31} \cdot \bm{b}_{32}) / m$.
We note that this cancellation is not specific to the P\"oschl-Teller
potential, but occurs generally if $\phi$ is taken to be the exact
zero-energy solution of the two-body Schr\"odinger equation with
$V(r_{ij})$.
The present formalism is therefore applicable to a broad class of
interaction potentials whose two-body Schr\"odinger equation can be
solved accurately, either analytically or numerically.
Furthermore, as shown in Sec.~\ref{s:fermi}, the above formalism
remains effective even when the exact two-body solution is
unavailable: an approximate $\phi$ can be used to construct
$V_{\rm eff}$ according to Eq.~(\ref{cal2}), despite the absence of
the exact cancellation.

In a previous work~\cite{boson1}, the ground states for several bosons
interacting through the Gaussian potential were calculated using
neural-network quantum states without the two-body correlations
$\varphi$.
As we will show in the next subsection, the inclusion of $\varphi$ is
crucial for increasing the accuracy and stability of the calculations
for practical use, which enables the study of the excited states.

\subsection{Results}
\label{s:bose_results}

We present numerical results for the ground and first excited states
of $N$-boson systems with $N=3$, 4, 5, and 6 interacting via the
unitary P\"oschl-Teller potential.
To assess the accuracy of the present method, we compare our results
with those obtained using the hyperspherical harmonic expansion
method~\cite{finite}.

\begin{figure}[t]
  \centering
  \includegraphics[width=0.98\linewidth]{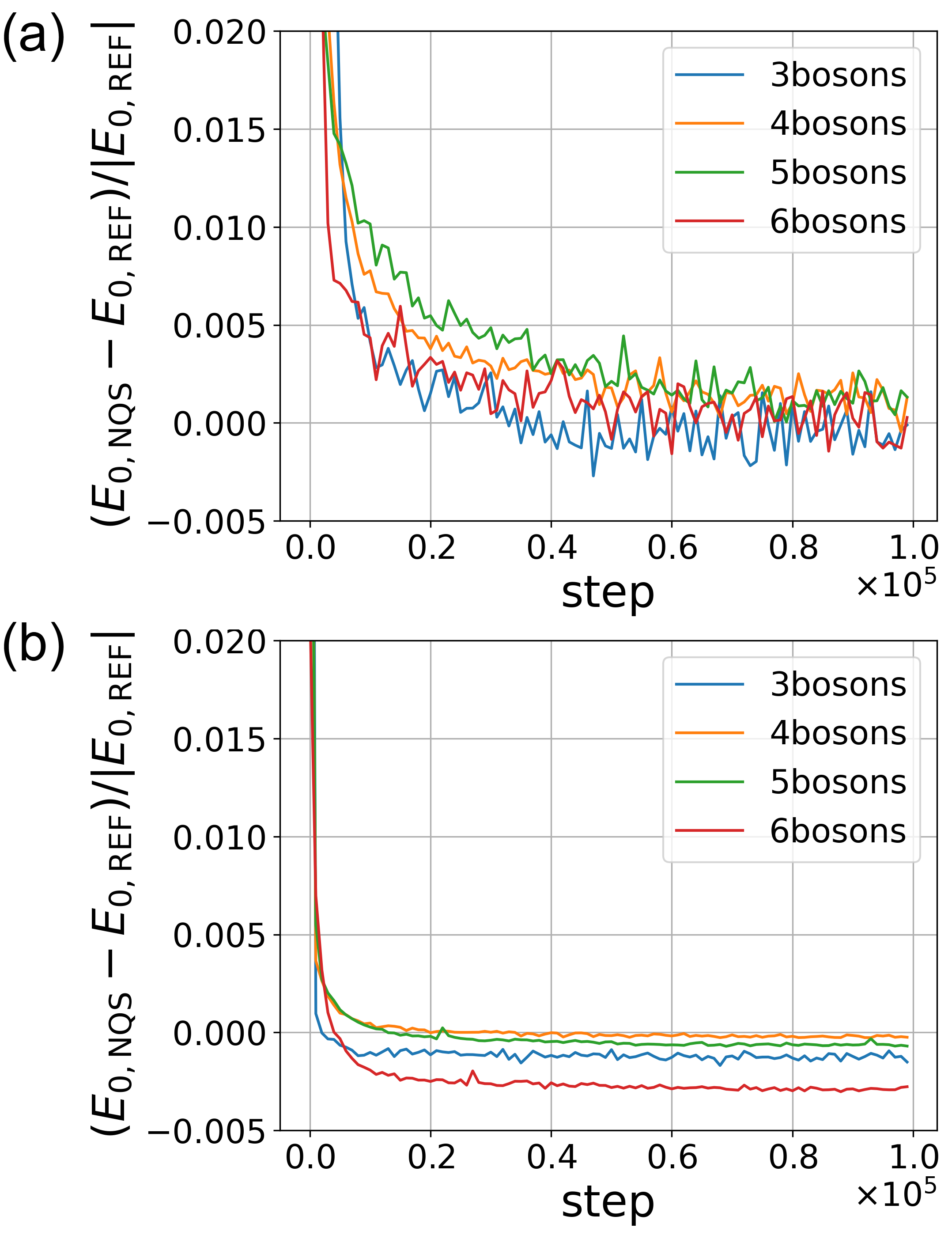}
  \caption{
    Optimization process of neural networks for $N$-boson ground
    states.
    The horizontal axis represents the number of optimization steps
    and the vertical axis represents the relative difference between
    the obtained ground-state energy $E_{0,\mathrm{NQS}}$ and the
    reference energy $E_{0,\mathrm{REF}}$ taken from
    Ref.~\cite{finite}.
    (a) Two-body solution is not included in the trial wave function:
    Eq.~(\ref{A}) is used as the wave function and the expectation
    value of the Hamiltonian in Eq.~(\ref{Hprime}) is evaluated.
    (b) Two-body solution is included in the trial wave function:
    the expression in Eq.~(\ref{finalH}) with Eq.~(\ref{Veff}) is
    evaluated with the wave function in Eqs.~(\ref{phiA}) and
    (\ref{varphi}) with Eq.~(\ref{sol}).
    In both cases, $10^{5}$ Monte Carlo samples are used in each step,
    and the values of the energies are averaged over every 1000 update
    steps.
  }
  \label{fig:ground}
\end{figure}

Figure~\ref{fig:ground} shows the convergence behavior of the
ground-state energy with respect to the update steps.
In Fig.~\ref{fig:ground}(a), the network output in Eq.~(\ref{A}) is
directly used as the wave function and the Hamiltonian in
Eq.~(\ref{Hprime}) is evaluated.
In Fig.~\ref{fig:ground}(b), the two-body wave function $\varphi$ is
used as in Eq.~(\ref{phiA}) and the energy in Eq.~(\ref{finalH}) is
evaluated.
For both methods, the $N$-body ground-state energies converge to the
values in Ref.~\cite{finite}, demonstrating that the neural networks
can accurately describe the $N$-body Borromean clusters associated
with the Efimov states.
A comparison between Figs.~\ref{fig:ground}(a) and \ref{fig:ground}(b)
clearly shows that the inclusion of the two-body correlation $\varphi$
significantly improves the stability and smoothness of the
convergence, while noticeable fluctuations are observed without
$\varphi$.
This improvement originates from the more accurate description of the
short-range two-body correlations by $\varphi$, by which the network
does not need to represent the sharp short-range profile of the wave
function.
Notably, for all $N$, the converged ground-state
energies in Fig.~\ref{fig:ground}(b) are slightly lower than the
reference values taken from Ref.~\cite{finite}.
Since the present approach is based on the variational principle,
the inequality $E_{0,\mathrm{NQS}} \ge E_{0}$ must hold, where $E_0$
is the exact ground-state energy. 
Therefore, the negative values obtained in Fig.~\ref{fig:ground}(b)
indicate that our method provides more accurate energies than those
in Ref.~\cite{finite}.

\begin{figure}[t]
  \centering
  \includegraphics[width=0.98\linewidth]{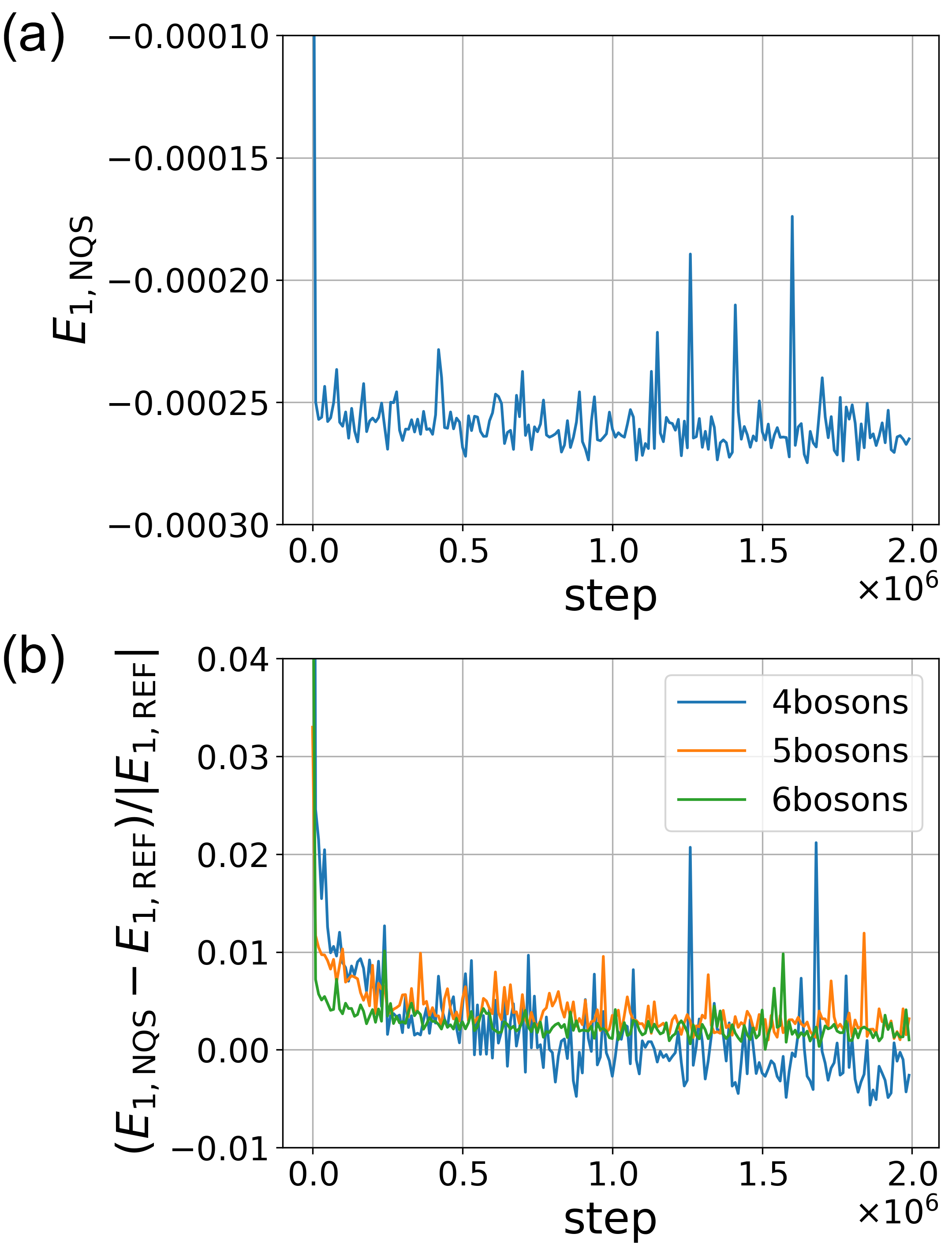}
  \caption{
    Optimization process of neural networks for first excited
    states of $N$-boson systems.
  (a) Energies for $N = 3$, normalized by $\hbar^2 / (mb^2)$.
  (b) Relative energy difference for $N = 4$, 5, and 6, where the
    values of $E_{1,\mathrm{REF}}$ are taken from Ref.~\cite{finite}.
    Since the excited-state energy for $N=3$ is not explicitly
    provided in Ref.~\cite{finite}, only $E_{1,\mathrm{NQS}}$ is
    plotted in (a).
    The values of the energies are averaged over every 10\,000 update
    steps.
  }
  \label{fig:excite}
\end{figure}

Using the obtained ground-state energies and wave functions,
we next compute the first excited state.
In the following, we only present the results obtained with the
two-body correlation $\varphi$.
To avoid the numerical instability originating from the small binding
energies and large spatial extent of excited few-body
clusters~\cite{von2009signatures, vonStecher2010, finite,
  PhysRevA.90.052514}, an auxiliary harmonic potential in
Eq.~(\ref{aux}) with $\omega = 10^{-3} \hbar / (m b^2)$ is applied
during the first 1000 steps.
Figure~\ref{fig:excite} shows the convergence behavior of the energies
of the first excited states.
For all $N$, the excited-state energies converge to negative values,
indicating the formation of bound states.
Although the fluctuations in the energies in Fig.~\ref{fig:excite}(b)
for $N = 4$--$6$ are $\sim 1\%$, those in Fig.~\ref{fig:excite}(a) for
$N = 3$ are $\sim 10\%$.
The large fluctuation for $N = 3$ is due to the extended spatial
structure of the Efimov state;
the spatial scale of the excited state is $\simeq 20$ times larger
than that of the ground state, while the short-range node structure is
also important for ensuring orthogonality with the ground state.
This multiscale property of the Efimov states makes the Monte Carlo
sampling less efficient.
For $N = 4$--$6$, by contrast, the first excited $N$-body state is tied
below the $(N-1)$-body ground
state~\cite{von2009signatures,RevModPhys.89.035006,naidon,d2018few}
and therefore remains relatively compact.
Consequently, the fluctuations arising from the Monte Carlo sampling
are less significant.

\begin{table}[t]
\centering
\caption{
Ground-state energies $E_0$, first excited-state energies $E_1$,
and ratios $\sqrt{E_0 / E_1}$ for $N$-boson systems.
The value of 22.4 for $N = 3$ is taken from the main text of
Ref.~\cite{finite} and the other reference values are taken from
Table~II in Ref.~\cite{finite}.
The energies are normalized by $\hbar^2 / (mb^2)$.
}
\label{tab:energy_scale}
\begin{tabular}{c|ccc}
\hline
 & $E_0$ & $E_1$ & $\sqrt{E_0 / E_1}$ \\ \hline
$N=3$ (REF) & $-0.1345$ &  & $22.4$ \\
$N=3$ (NQS) & $-0.1347$ & $-2.63\times10^{-4}$ & $22.6$ \\ \hline
$N=4$ (REF) & $-0.8259$ & $-0.1548$ & $2.31$ \\
$N=4$ (NQS) & $-0.8261$ & $-0.1550$ & $2.31$ \\ \hline
$N=5$ (REF) & $-2.313$ & $-0.9428$ & $1.57$ \\
$N=5$ (NQS) & $-2.315$ & $-0.9399$ & $1.57$ \\ \hline
$N=6$ (REF) & $-4.731$ & $-2.667$ & $1.33$ \\
$N=6$ (NQS) & $-4.745$ & $-2.663$ & $1.33$ \\ \hline
\end{tabular}
\end{table}

Using the trained neural networks, we next determine the energies of
the ground and first excited states.
In this post-trained evaluation, each energy is obtained from $10^{8}$
Monte Carlo samples, followed by 100 additional parameter updates of
the neural network.
This procedure is repeated 1000 times and the average of the energies
is taken.
The results are summarized in Table~\ref{tab:energy_scale}, where they
are compared with the values in Ref.~\cite{finite}.
Our values are in good agreement with those in Ref.~\cite{finite}.
For $N = 3$, the ratio $\sqrt{E_0 / E_1}$ slightly deviates from the
universal value predicted from the zero-range theory, $\simeq 22.7$,
which is attributed to finite-range effects of the P\"oschl-Teller
potential.

\begin{figure}[t]
  \centering
  \includegraphics[width=0.98\linewidth]{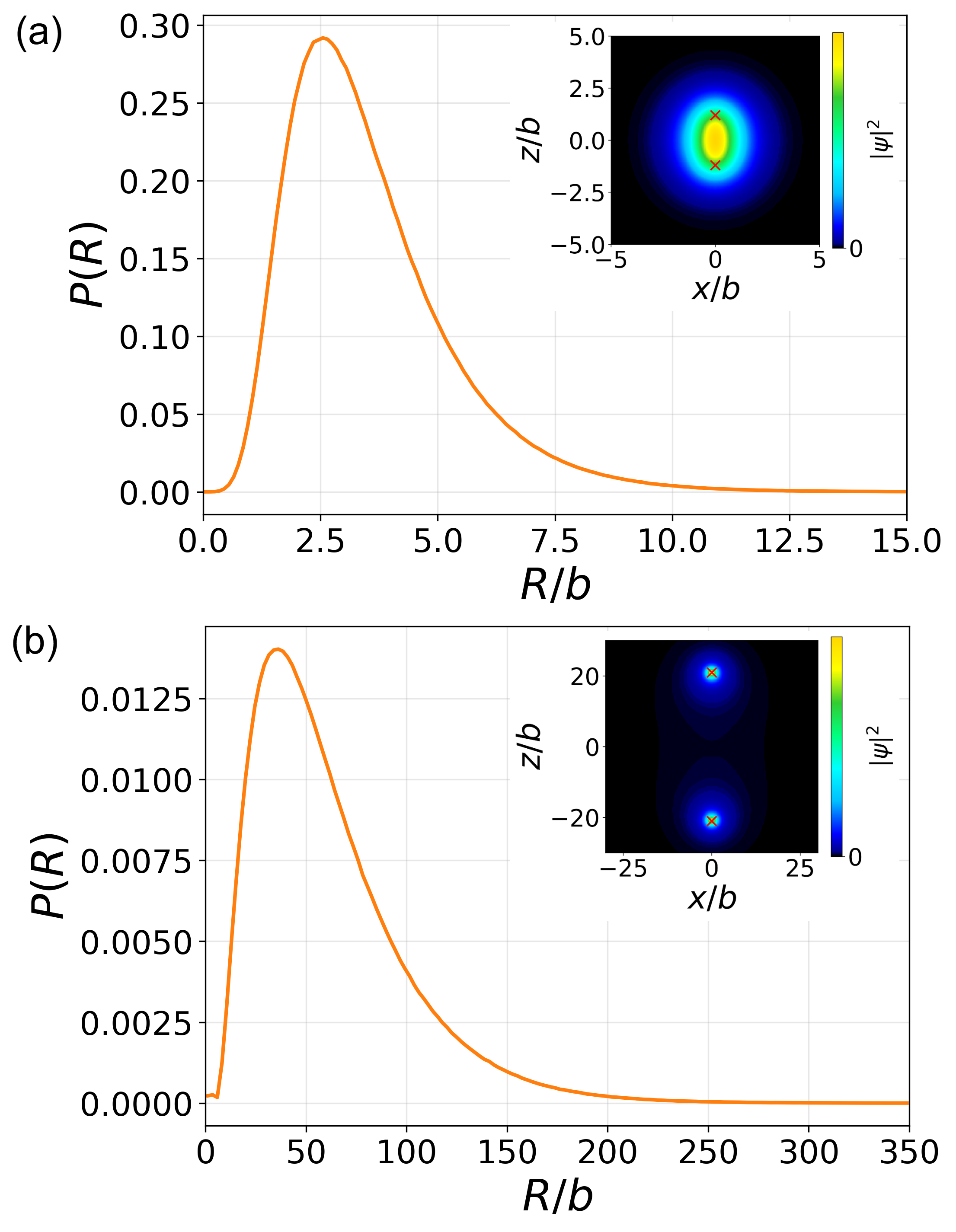}
  \caption{\label{fig:FR_boson}
    Probability distribution $P(R)$ as function of hyperradius
    $R=\sqrt{\rho_1^2+\rho_2^2}$ for bosonic three-body system:
    (a) ground state and (b) first excited state.
    $P(R)$ is generated from the wave function $\Psi$ as a histogram
    of $R$ with $10^7$ samples.
    The insets show the probability distribution for the position of
    the third particle when the other two particles are fixed at the
    positions indicated by the crosses.
  }
\end{figure}

Figures~\ref{fig:FR_boson}(a) and \ref{fig:FR_boson}(b) show the
probability distributions $P(R)$ for the ground and first excited
states, respectively, of the Efimov system ($N=3$).
Here, $P(R)$ is obtained as a histogram of the hyperradius $R =
\sqrt{\rho_1^2 + \rho_2^2}$, which characterizes the spatial extent of
the three-body system.
In both states, the distribution $P(R)$ exhibits a pronounced peak at
finite $R$ and decays exponentially at large $R$, indicating the
formation of bound states.
The two distributions have a similar shape with a scale ratio of about
20, consistent with the discrete scale invariance of the Efimov
states.
In the first excited state in Fig.~\ref{fig:FR_boson}(b), a node
appears in the short-range region, making the state orthogonal to the
ground state.

The insets in Fig.~\ref{fig:FR_boson} show the spatial distribution of
the three-body wave function, namely the probability distribution for
the position of the third particle when the other two particles are
fixed at the positions denoted by the crosses.
The results show the same qualitative features as those observed for
$^4$He trimers~\cite{Kunitski}: in contrast to the probability
density for the ground state [Fig.~\ref{fig:FR_boson}(a)], which
forms a compact bonding-like orbital, that for the first excited state
[Fig.~\ref{fig:FR_boson}(b)] is concentrated around each particle,
reflecting the origin of the Efimov trimer as arising from the
exchange of a particle between the other two particles in a
classically forbidden region.


\section{System of two identical fermions and one particle}
\label{s:fermi}

In this section, we apply the neural-network method to a three-body
system composed of two identical fermions (mass $M$) and a third
particle (mass $m$) interacting via unitary inter-species
interactions.
While the equal-mass system does not exhibit any three-body bound
states due to the Pauli repulsion between the identical fermions, it is
dominated by the attraction mediated by the third particle if the mass
ratio exceeds the critical value,
$M/m>13.606\ldots$~\cite{efimov1973energy, PhysRevA.67.010703, Blume,
  Daily}.
Such highly mass-imbalanced fermionic systems have recently been
realized in cold atom mixtures of Er-Li~\cite{er} and
Dy-Li~\cite{DyLifeshbach}, where the fermionic Efimov states are
expected to appear in the $\ell^{\Pi}=1^-$ angular-momentum and parity
sector~\cite{OiEndo2024,oi2025universal}.

\subsection{\label{s:fermi_method}Method}

We consider a three-body system composed of two identical fermions
located at $\bm{r}_1$ and $\bm{r}_2$ with mass $M$, and another
particle located at $\bm{r}_3$ with mass $m$.
The two-body interaction is introduced only between each fermion and
the third particle; that is, no interaction is assumed between the two
identical fermions, since there is no $s$-wave interaction between
them.
The Hamiltonian for this system is given by
\begin{eqnarray} \label{HF}
\hat{H}_F & = & -\frac{\hbar^2}{2M}
\left( \frac{\partial^2}{\partial\bm{r}_1^2}
+ \frac{\partial^2}{\partial\bm{r}_2^2}
\right)
- \frac{\hbar^2}{2m} \frac{\partial^2}{\partial\bm{r}_3^2}
\nonumber \\
& & + V\left( r_{13} \right) + V\left( r_{23} \right).
\end{eqnarray}
Using the Jacobi coordinates,
\begin{subequations} \label{JacobiF}
\begin{eqnarray}
\bm{\rho}_1 & = &
\sqrt{\frac{M + m}{2m}}
\left(
\bm{r}_2 - \bm{r}_1
\right),
 \\
\bm{\rho}_2 & = &
\sqrt{\frac{2(M + m)}{2M + m}}
\left(
\bm{r}_3 - \frac{\bm{r}_1+\bm{r}_2}{2}
\right),
\end{eqnarray}
\end{subequations}
we can eliminate the center-of-mass motion from the Hamiltonian and obtain 
\begin{equation} \label{HF2}
\hat{H}_F' =
-\frac{\hbar^2}{2\mu} \frac{\partial^2}{\partial\bm{\rho}_1^2}
-\frac{\hbar^2}{2\mu} \frac{\partial^2}{\partial\bm{\rho}_2^2}
+ V_{\mathrm{PT}}\left( r_{13} \right)
+ V_{\mathrm{PT}}\left( r_{23} \right),
\end{equation}
where $\mu = Mm / (M+m)$ is the reduced mass,
$r_{13} = |\sqrt{m} \bm{\rho}_1 + \sqrt{2M+m} \bm{\rho}_2| /
\sqrt{2(M + m)}$, and $r_{23} = |\sqrt{m} \bm{\rho}_1 - \sqrt{2M+m}
\bm{\rho}_2| / \sqrt{2(M + m)}$.

The Hamiltonian in Eq.~(\ref{HF2}) commutes with the angular-momentum
operator,
\begin{equation}
\hat{L} = \sum_{j=1}^2 \left(
\bm{\rho}_j \times \frac{\hbar}{i}
\frac{\partial}{\partial \bm{\rho}_j} \right).
\end{equation}
The eigenstates of the Hamiltonian $\hat{H}_F'$ can therefore be
characterized by the angular-momentum quantum numbers.
We express the Jacobi vectors $\bm{\rho}_1$ and $\bm{\rho}_2$ by
their lengths $\rho_1$ and $\rho_2$, the angle $\theta$ between them,
and their overall rotation by the Euler angles $\phi$, $\theta'$, and
$\phi'$ as
\begin{equation} \label{rho1}
  \bm{\rho}_1 = R_z(\phi') R_y(\theta') R_z(\phi) \left(
  \begin{array}{c} 0 \\ 0 \\ \rho_1 \end{array} \right)
  = \rho_1 \left(\begin{array}{c} \sin\theta' \cos\phi' \\ \sin\theta'
    \sin\phi' \\ \cos\theta' \end{array} \right),
\end{equation}
\begin{widetext}
\begin{equation} \label{rho2}
    \bm{\rho}_2 = R_z(\phi') R_y(\theta') R_z(\phi) \left(
  \begin{array}{c} \rho_2 \sin\theta \\ 0 \\ \rho_2
    \cos\theta \end{array} \right)
  = \rho_2 \left( \begin{array}{c}
    \cos\theta \sin\theta' \cos\phi' + \sin\theta
    (\cos\phi \cos\theta' \cos\phi' - \sin\phi \sin\phi') \\
    \cos\theta \sin\theta' \sin\phi' + \sin\theta
    (\cos\phi \cos\theta' \sin\phi' + \sin\phi \cos\phi') \\
    \cos\theta \cos\theta' - \sin\theta \cos\phi \sin\theta' 
  \end{array} \right),
\end{equation}
\end{widetext}
where $R_{x, y, z}(\alpha)$ is the rotation matrix, with rotation by
an angle $\alpha$ around the $x$, $y$, or $z$ axis.
In this coordinate system, the operator $\hat{\bm{L}}^2$ can be
represented only by the Euler angles (it does not include $\rho_1$,
$\rho_2$, and $\theta$)~\cite{breit}.
We denote the eigenvalues of $\hat{\bm{L}}^2$ and $\hat{L}_z =
\hbar \partial / (i \partial\phi')$ as $\hbar^2 \ell (\ell + 1)$ and
$\hbar m_z$, respectively, and the corresponding eigenfunction as
$Y_{n, \ell, m_z}(\phi, \theta', \phi')$, where the index $n$
identifies the eigenstates within each angular-momentum subspace.
The energy is degenerate with respect to $m_z$ due to the rotational
symmetry and we only consider the $m_z = 0$ states.
We numerically confirmed that the $\ell = 1$ subspace with odd
parity has the lowest energy, consistent with the zero-range results,
and therefore focus on this channel throughout this section.
We note, however, that our formalism can be systematically extended to
other angular-momentum and parity channels.
The parity transformation,
$\bm{\rho}_1 \rightarrow -\bm{\rho}_1$
and $\bm{\rho}_2 \rightarrow -\bm{\rho}_2$, is rewritten as
$\theta' \rightarrow \pi - \theta'$, $\phi \rightarrow \pi - \phi$,
and $\phi' \rightarrow \phi' + \pi$.
There are nine eigenfunctions with $\ell = 1$~\cite{breit}.
Among them, the eigenfunctions with $m_z = 0$ and odd parity are given
by
\begin{equation}
Y_{1, 1, 0} = \cos\theta', \qquad
Y_{2, 1, 0} = \sin\theta'\cos\phi.
\end{equation}
The general form of the wave function can therefore be written as
\begin{equation}
\begin{aligned}
\psi(\rho_1,\rho_2,\theta;\theta',\phi)
={}&
f_1(\rho_1,\rho_2,\theta)\cos\theta'
\\
&+ f_2(\rho_1,\rho_2,\theta)\sin\theta'\cos\phi,
\end{aligned}
\end{equation}
where $f_1$ and $f_2$ are unknown functions.

We express the functions $f_1$ and $f_2$ by two different neural
networks as $f_1 = A(P, W_1)$ and $f_2 = A(P, W_2)$, respectively,
where $A$ is the network output given in Eq.~(\ref{A}).
The wave function must be antisymmetric with respect to the exchange
of the two identical fermions.
In the Jacobi coordinates, the exchange of the fermions is expressed
as $\bm{\rho}_1 \rightarrow -\bm{\rho}_1$ and
$\bm{\rho}_2 \rightarrow \bm{\rho}_2$, which are rewritten as
$\theta \rightarrow \pi - \theta$,
$\theta' \rightarrow \pi - \theta'$, and $\phi \rightarrow -\phi$, as
found from Eqs.~(\ref{rho1}) and (\ref{rho2}).
Thus, in a manner similar to Eqs.~(\ref{phiA}) and (\ref{varphi}), we
construct the trial wave function as
\begin{equation} \label{phiasym}
  \Psi = \varphi A_{\rm asym},
\end{equation}
where
\begin{equation} \label{phi2}
\varphi = \phi(r_{13}) \phi(r_{23})
\end{equation}
is the product of two-body correlations and
\begin{equation}
  A_{\rm asym} = \psi(\rho_1, \rho_2, \theta, \theta', \phi)
- \psi(\rho_1, \rho_2, \pi-\theta, \pi-\theta', -\phi)
\end{equation}
ensures antisymmetrization for the fermions.
As in the bosonic case in Eq.~(\ref{cal2}), we can define the
effective potential $V_{\rm eff}$ arising from the derivative of
$\varphi$.
The energy integral is written as
\begin{equation}
  \int \Psi \hat{H}_F' \Psi dP =
\int dP \left[ \frac{\hbar^2}{2\mu} \sum_{j=1}^{2}
  \varphi^2 \left( \frac{\partial A_{\rm asym}}{\partial\bm{\rho}_j}
  \right)^2 + V_{\rm eff} \Psi^2 \right],
\end{equation}
where $dP = d\bm{\rho}_1 d\bm{\rho}_2$.
The procedures used to optimize the network parameters are the same as
those for the bosonic case.
The first excited state is searched within the same subspace as
the ground state, namely the $\ell = 1$ and odd parity subspace.

In the following numerical calculations, we use the P\"oschl-Teller
potential for the two-body interaction potential as
\begin{equation} \label{PT}
  V_{\mathrm{PT}}(r) = -\frac{\hbar^2}{\mu b^2}
  \frac{1}{\cosh^2 \frac{r}{b}},
\end{equation}
whose $s$-wave scattering length is infinite, corresponding to the
unitary limit.
We use the zero-energy analytical solution of this potential in
Eq.~(\ref{sol}) as the two-body correlation $\phi$ in Eq.~(\ref{phi2}).
The two-body potential term in the Hamiltonian vanishes, as in the
bosonic case, and the effective potential simplifies to
\begin{equation} \label{fVeff}
  V_{\rm eff} = -\frac{\hbar^2}{m} \bm{b}_{13} \cdot \bm{b}_{23},
\end{equation}
where $\bm{b}_{ij}$ is defined in Eq.~(\ref{b2}).

\subsection{\label{s:fermi_results}Results}

First, we focus on the mass ratio $M / m = 27.752$ corresponding to the
$^{167}\mathrm{Er}$-$^{6}\mathrm{Li}$ system.
A mixture of these ultracold atomic gases has been realized
experimentally, in which interspecies Feshbach resonances were
observed~\cite{er}.
By fine-tuning the $s$-wave scattering length to be large and achieving
the unitary limit, the Efimov states involving identical fermions are
expected to appear in the $\ell=1$ channel when the mass ratio is
larger than the critical value, $M/m \gtrsim 13.606$.
While a strong dipole-dipole interaction between the Er atoms may
affect the physical properties of the Efimov states of Er-Er-Li, we
neglect this interaction in this work and only consider the
short-range interactions between Er and Li, tuned to the unitary
limit, and demonstrate that our neural-network method is useful for
investigating fermionic Efimov physics.

\begin{figure}[t]
  \centering
  \includegraphics[width=0.9\linewidth]{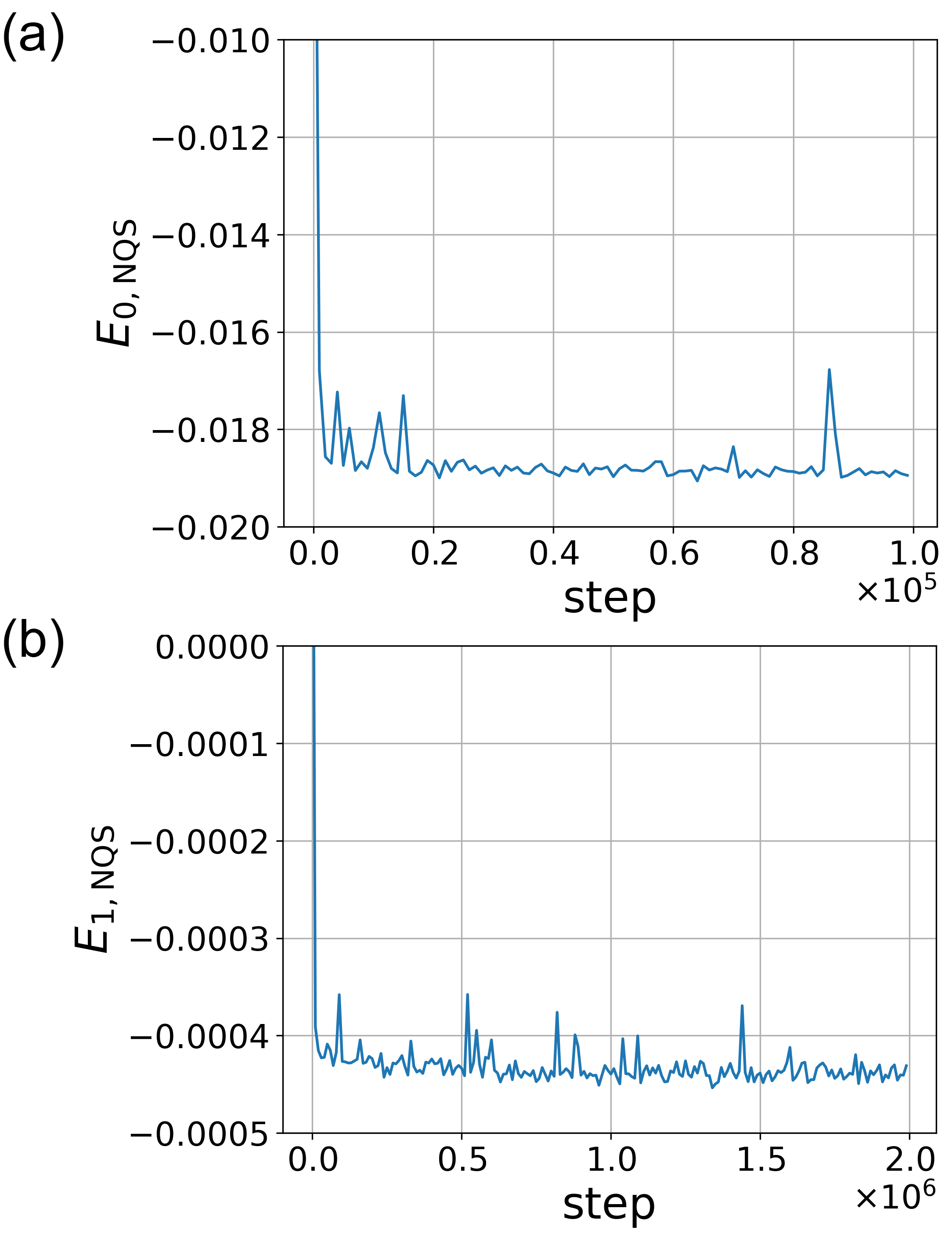}
  \caption{
    Optimization process of neural networks for system of two
    identical fermions and another particle with mass ratio $M / m =
    27.752$.
    (a) Ground-state energy and (b) first excited-state energy,
    normalized by $\hbar^2 / (m b^2)$.
    $10^{5}$ Monte Carlo samples are taken in each update.
    Auxiliary harmonic potential in Eq.~(\ref{aux}) is applied during
    the first 1000 steps, with $\omega=10^{-2} \hbar / (m b^2)$ for
    the ground state and $\omega=10^{-3} \hbar / (m b^2)$ for the
    excited state.
    The optimization well converges for $\sim 10^{5}$ steps in (a),
    while $\sim 10^{6}$ steps are needed for (b).
  }
  \label{fig:fermi}
\end{figure}

Figure~\ref{fig:fermi} shows the convergence of the energies with
respect to the network parameters' update steps.
For both the ground and first excited states, the energies converge to
negative values, indicating the formation of bound states.
The excited state exhibits larger statistical fluctuations than those
for the ground state, reflecting its extended spatial structure, which
makes the Monte Carlo integration more difficult, as in the bosonic
case.
After the convergence of the optimization, we estimate the energies
with $10^8 \times 10^3$ Monte Carlo samples, in a manner similar to
the bosonic case, giving $E_0 = -1.880 \times 10^{-2} \hbar^2 / (m
b^2)$ for the ground state and $E_1 = -4.37 \times 10^{-4} \hbar^2 /
(m b^2)$ for the first excited state.
The ratio $\sqrt{|E_0/E_1|} \simeq 6.6$ is in
reasonable agreement with the universal discrete scale factor
predicted from the zero-range theory,
$\sqrt{|E_0/E_1|}\simeq 7.19$~\cite{efimov1973energy, Braaten2006259,
  RevModPhys.89.035006, naidon, d2018few, PhysRevA.67.010703}
for $M / m = 27.752$.
The small deviation is attributed to the finite-range effects of the
potential and the limited validity of the low-energy condition, which
often deteriorate the universal description of the ground Efimov
trimer.

\begin{figure}[t]
  \centering
  \includegraphics[width=0.98\linewidth]{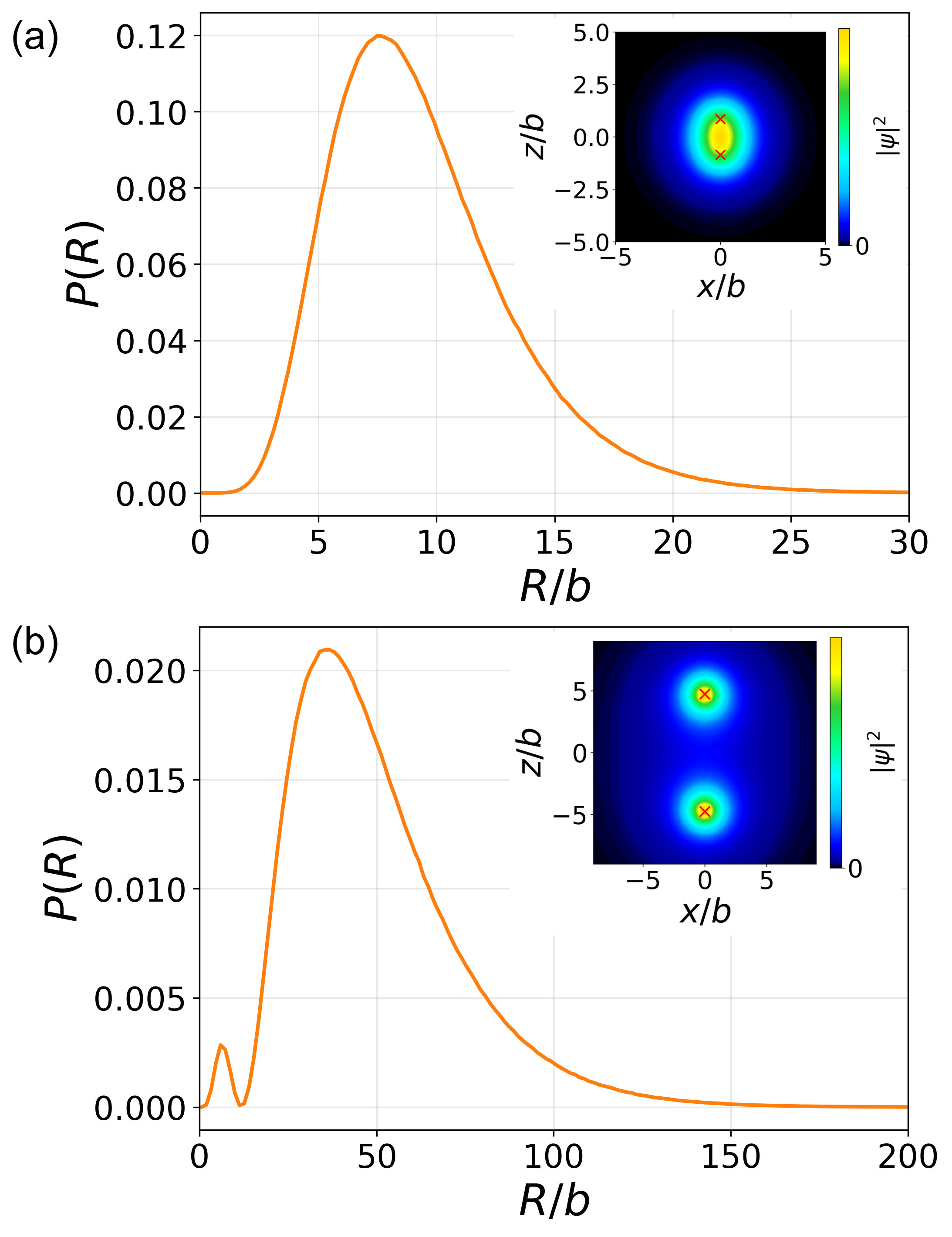}
  \caption{
    Probability distribution $P(R)$ as function of hyperradius
    $R$ for system of two identical fermions and one particle with
    mass ratio $M / m = 27.752$: (a) ground state and (b) first
    excited state.
    The insets show the probability distributions for the position of
    the third particle when the two identical fermions are fixed at
    the positions indicated by the crosses.
  }
  \label{fig:FR_fermi}
\end{figure}

Figures~\ref{fig:FR_fermi}(a) and \ref{fig:FR_fermi}(b) show the
probability distributions $P(R)$ with respect to the hyperradius $R$
for the ground and first excited states, respectively.
As in the bosonic case, the two distributions have similar shapes with
different spatial scales, exhibiting the discrete scale invariance in
Efimov physics.
The scale ratio is $\simeq 5$, which is consistent with the zero-range
prediction of $\simeq 7.19$.
In the first excited state [Fig.~\ref{fig:FR_fermi}(b)], a node
appears in the short-range region, makeing the state orthogonal to the
ground state.
The insets in Fig.~\ref{fig:FR_fermi} show the probability
distribution for the position of the third particle when the two
identical fermions are fixed.
The results show the same qualitative features as those for the
bosonic three-body systems in Fig.~\ref{fig:FR_boson}, demonstrating
that the neural network can capture the spatial structure of fermionic
Efimov states.

\begin{figure}[t]
  \centering
  \includegraphics[width=0.98\linewidth]{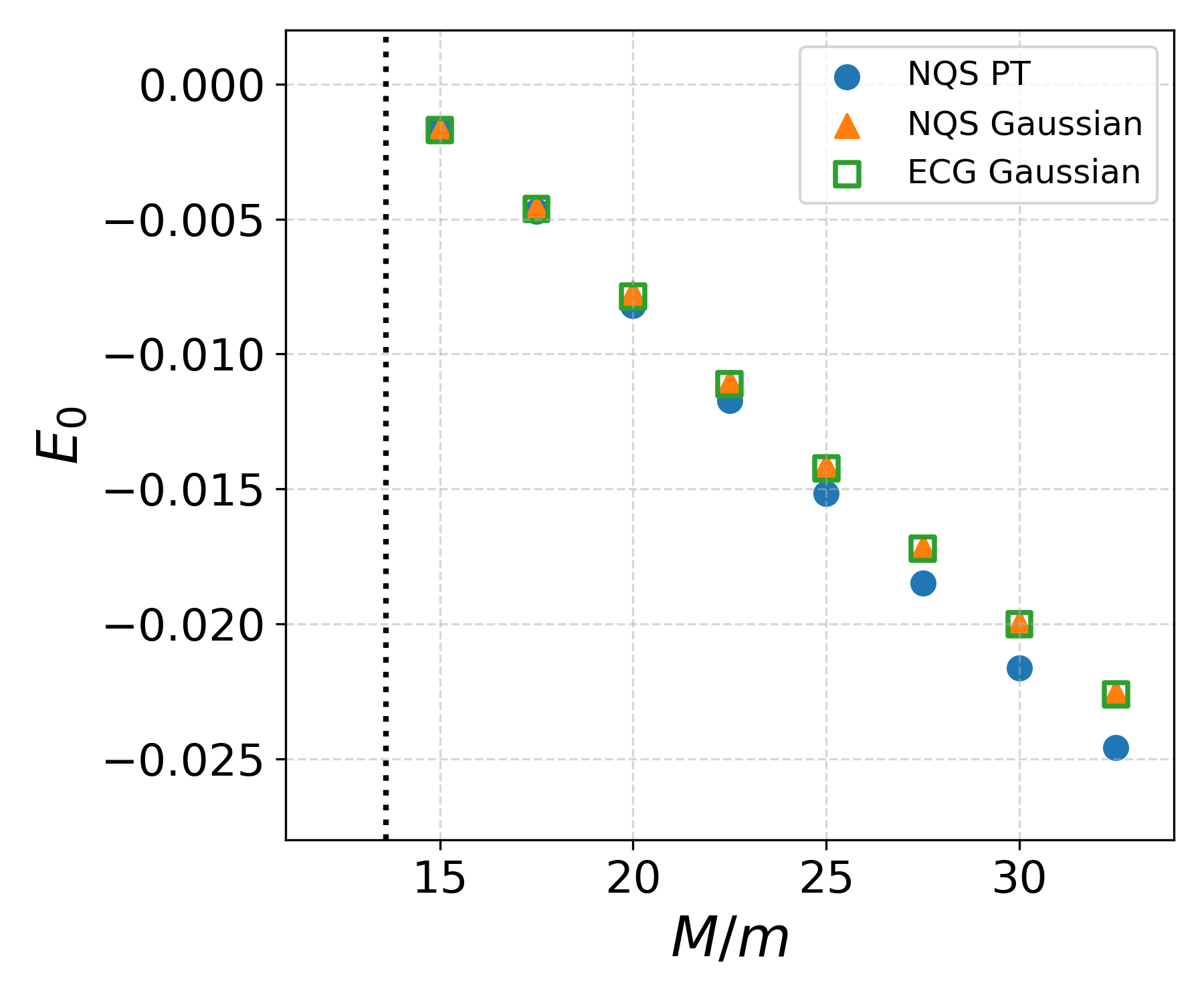}
  \caption{
  Ground-state energy [normalized by $\hbar^2 / (m b^2)$] of
  system of two identical fermions and one particle as function of
  mass ratio $M / m$.
  Circles show the results obtained using the neural-network method
  for the P\"oschl-Teller potential at the unitary limit in
  Eq.~(\ref{PT}).
  Triangles show the results obtained using the neural-network method
  for the Gaussian potential in Eq.~(\ref{VG}) at the unitary limit.
  Squares show the results obtained using the explicitly-correlated
  Gaussian-basis method for the same Gaussian
  potential~\cite{Blume, Daily}. 
  The vertical dotted line indicates $M / m = 13.606$.
  }
  \label{f:mass}
\end{figure}

We also perform the calculation for variable mass ratios and obtain
the dependence of the ground-state energy on the mass ratio $M / m$,
as shown by the circles in Fig.~\ref{f:mass}.
As expected from the zero-range Efimov theory~\cite{efimov1973energy,
  Braaten2006259, RevModPhys.89.035006, naidon, d2018few},
the binding energy increases monotonically with $M/m$.
The data extrapolate to the disappearance point of the bound state,
consistent with the critical mass ratio of
$13.606\ldots$~\cite{efimov1973energy, PhysRevA.67.010703}, confirming
that the neural-network method can reliably describe fermionic
few-body systems even near the binding threshold.

We finally consider the case in which the two-body interaction
potential is not a P\"oschl-Teller potential.
We consider the Gaussian potential
\begin{equation} \label{VG}
  V_G(r_{ij}) = V_0 e^{-r_{ij}^2 / b^2},
\end{equation}
where $b$ characterizes the interaction range.
The value of $V_0$ is taken to be $-1.342 \hbar^2 / (\mu b^2)$,
corresponding to the depth at which the two-body ground bound state is
about to appear and hence to the unitary limit.
Although the analytical two-body solution is unavailable for the
Gaussian potential, we can still perform the neural-network
calculation by using the two-body correlation of the P\"oschl-Teller
potential, exploiting the similar, though not identical, short-range
correlations of the two potentials.
More specifically, in solving the problem with the Gaussian potential,
we use the trial wave function in Eqs.~(\ref{phiasym}) and
(\ref{phi2}) with the P\"oschl-Teller two-body solution in
Eq.~(\ref{sol}).
The effective potential changes from Eq.~(\ref{fVeff}) to
\begin{eqnarray}
  V_{\rm eff} & = & -\frac{\hbar^2}{m} \bm{b}_{13} \cdot \bm{b}_{23}
  + V_G(r_{13}) + V_G(r_{23})
  \nonumber \\
  & & - V_{\rm PT}(r_{13}) - V_{\rm PT}(r_{23}).
\end{eqnarray}
The obtained ground-state energies are shown by the
triangles in Fig.~\ref{f:mass}.
They are in excellent agreement with those obtained using the
explicitly-correlated Gaussian-basis method~\cite{Blume, Daily} for
the same two-body Gaussian potential (squares in Fig.~\ref{f:mass}).
This agreement is attributed to the ability of the neural network to
compensate for the relatively small differences between the
short-range two-body correlations of the P\"oschl-Teller and Gaussian
potentials, thereby yielding accurate energies.
The above results suggest that the neural-network method is not
limited to analytically solvable two-body potentials, but is
applicable to problems with a wide variety of interaction potentials.

\section{Conclusions and Discussion}
\label{s:conc}
We developed the method of neural-network quantum states for
exploring Efimov physics in few-body quantum systems in
three-dimensional continuous space.
By using the two-body solution to efficiently incorporate the strong
two-body correlations at the unitary limit into the variational wave
function, the convergence and accuracy are significantly improved.
For systems of identical $N$ bosons, as well as the
mass-imbalanced three-body fermionic system, the ground and first
excited states are obtained with an accuracy comparable to or better
than previously reported results.
The ratio between the ground and first excited-state energies in the
three-body systems agrees well with the values predicted by the
universal Efimov theory.
In addition, our method reproduces the characteristic shapes of the
Efimov states' wave functions and captures the disappearance of the
Efimov state as the mass ratio approaches the critical value of
$13.606\ldots$ for mass-imbalanced fermions.
These results demonstrate that the neural-network quantum states
provide a versatile and accurate approach for exploring few-body
systems near the unitary limit.

In most of the numerical calculations presented in this paper, we used
the P\"oschl-Teller potential, for which the analytical form of the
two-body correlation $\phi$ and its derivative $\bm{b}_{ij}$ are
available, with which the convergence is accelerated.
However, the proposed method is not restricted to analytically
solvable potentials.
Indeed, leveraging the fact that the two-body correlations $\phi$
exhibit similar, though not identical, short-range behavior, we used
the two-body correlation of the P\"oschl-Teller potential to study a
three-body system with a Gaussian potential.
The neural networks are found to compensate for the small differences
in the two-body correlations of different potentials and to converge
rather efficiently to the accurate values obtained using an
established method~\cite{Blume, Daily}.
Alternatively, one may employ the numerically obtained $\phi$ and
$\bm{b}_{ij}$ with appropriate interpolation.
These schemes would allow solving few-body problems with general
interaction potentials, including regimes away from the unitary limit
considered here, or dipolar types of interactions whose long-range
and anisotropic nature leads to exotic few-body
clusters~\cite{PhysRevLett.106.233201, PhysRevLett.107.233201,
  OiEndo2024, oi2025universal, hcwf-tk6c, Ohishi}.

\begin{acknowledgments}
We thank D. Blume for providing the data shown as squares in
Fig.~\ref{f:mass}, which were calculated using the same methods as
used in Refs.~\cite{Blume, Daily}.
HS was supported by JSPS KAKENHI Grant Numbers JP23K03276 and
JP26K00638.
SE acknowledges support from Matsuo Foundation and JSPS KAKENHI Grant
Number JP25K00217.
\end{acknowledgments}


\begin{thebibliography}{84}%
\makeatletter
\providecommand \@ifxundefined [1]{%
 \@ifx{#1\undefined}
}%
\providecommand \@ifnum [1]{%
 \ifnum #1\expandafter \@firstoftwo
 \else \expandafter \@secondoftwo
 \fi
}%
\providecommand \@ifx [1]{%
 \ifx #1\expandafter \@firstoftwo
 \else \expandafter \@secondoftwo
 \fi
}%
\providecommand \natexlab [1]{#1}%
\providecommand \enquote  [1]{``#1''}%
\providecommand \bibnamefont  [1]{#1}%
\providecommand \bibfnamefont [1]{#1}%
\providecommand \citenamefont [1]{#1}%
\providecommand \href@noop [0]{\@secondoftwo}%
\providecommand \href [0]{\begingroup \@sanitize@url \@href}%
\providecommand \@href[1]{\@@startlink{#1}\@@href}%
\providecommand \@@href[1]{\endgroup#1\@@endlink}%
\providecommand \@sanitize@url [0]{\catcode `\\12\catcode `\$12\catcode
  `\&12\catcode `\#12\catcode `\^12\catcode `\_12\catcode `\%12\relax}%
\providecommand \@@startlink[1]{}%
\providecommand \@@endlink[0]{}%
\providecommand \url  [0]{\begingroup\@sanitize@url \@url }%
\providecommand \@url [1]{\endgroup\@href {#1}{\urlprefix }}%
\providecommand \urlprefix  [0]{URL }%
\providecommand \Eprint [0]{\href }%
\providecommand \doibase [0]{https://doi.org/}%
\providecommand \selectlanguage [0]{\@gobble}%
\providecommand \bibinfo  [0]{\@secondoftwo}%
\providecommand \bibfield  [0]{\@secondoftwo}%
\providecommand \translation [1]{[#1]}%
\providecommand \BibitemOpen [0]{}%
\providecommand \bibitemStop [0]{}%
\providecommand \bibitemNoStop [0]{.\EOS\space}%
\providecommand \EOS [0]{\spacefactor3000\relax}%
\providecommand \BibitemShut  [1]{\csname bibitem#1\endcsname}%
\let\auto@bib@innerbib\@empty
\bibitem{QMC}
  W. M. C. Foulkes, L. Mitas, R. J. Needs, and G. Rajagopal,
  Quantum Monte Carlo simulations of solids,
  Rev. Mod. Phys. \textbf{73}, 33 (2001). 

\bibitem{DMRG}
  U. Schollw\"ock,
  The density-matrix renormalization group,
  Rev. Mod. Phys. \textbf{77}, 259 (2005).

\bibitem{TN}
S.-J.~Ran, E.~Tirrito, C.~Peng, X.~Chen, L.~Tagliacozzo, G.~Su, and
M.~Lewenstein,
\textit{Tensor Network Contractions: Methods and Applications to
  Quantum Many-Body Systems},
Lecture Notes in Physics Vol. 964 (Springer International Publishing,
Cham, 2020).
 
\bibitem{NQS}
  G. Carleo and M. Troyer,
  Solving the Quantum Many-Body Problem with Artificial Neural Networks,
  Science \textbf{355}, 602 (2017).

\bibitem{review1}
  Z.-A. Jia, B. Yi, R. Zhai, Y.-C. Wu, G.-C. Guo, and G.-P. Guo,
  Quantum Neural Network States: A Brief Reviewof Methods and
  Applications,
  Adv. Quantum Technol. \textbf{2}, 1800077 (2019).

\bibitem{review2}
  R. G. Melko, G. Carleo, J. Carrasquilla, and J. I. Cirac,
  Restricted Boltzmann machines in quantum physics,
  Nat. Phys. \textbf{15}, 887 (2019).

\bibitem{review3}
  M. Medvidovi\'c and J. R. Moreno,
  Neural-network quantum states for many-body physics,
  Eur. Phys. J. Plus \textbf{139}, 631 (2024).

\bibitem{review4}
  H. Lange, A. Van de Walle, A. Abedinnia, and A. Bohrdt,
  From architectures to applications: a review of neural quantum states,
  Quantum Sci. Technol. \textbf{9}, 040501 (2024).

\bibitem{BHM}
  H. Saito,
  Solving the Bose-Hubbard Model with Machine Learning,
  J. Phys. Soc. Jpn. \textbf{86}, 093001 (2017).

\bibitem{Nomura1}
  Y. Nomura, A. S. Darmawan, Y. Yamaji, and M. Imada,
  Restricted Boltzmann machine learning for solving strongly
  correlated quantum systems,
  Phys. Rev. B \textbf{96}, 205152 (2017).

\bibitem{Cai}
  Z. Cai and J. Liu,
  Approximating quantum many-body wave functions using artificial
  neural networks,
  Phys. Rev. B \textbf{97}, 035116 (2018).

\bibitem{BHM2}
  H. Saito and M. Kato,
  Machine Learning Technique to Find Quantum Many-Body Ground States
  of Bosons on a Lattice,
  J. Phys. Soc. Jpn. \textbf{87}, 014001 (2018).

\bibitem{boson1}
  H. Saito,
  Method to solve quantum few-body problems with artificial neural networks,
  J. Phys. Soc. Jpn. \textbf{87}, 074002 (2018).

\bibitem{boson2}
  G. Pescia, J. Han, A. Lovato, J. Lu, and G. Carleo,
  Neural-network quantum states for periodic systems in continuous space,
  Phys. Rev. Res. \textbf{4}, 023138 (2022).
 
\bibitem{boson3}
  J. M. Martyn, K. Najafi, and D. Luo,
  Variational Neural-Network Ansatz for Continuum Quantum Field
  Theory,
  Phys. Rev. Lett. \textbf{131}, 081601 (2023).

\bibitem{boson4}
  T. Naito, H. Naito, and K. Hashimoto,
  Multi-body wave function of ground and low-lying excited states
  using unornamented deep neural networks,
  Phys. Rev. Res. \textbf{5}, 033189 (2023).

\bibitem{boson5}
  P. F. Bedaque, H. Kumar, and A. Sheng,
  Neural network solutions of bosonic quantum systems in one dimension,
  Phys. Rev. C \textbf{109}, 034004 (2024).

\bibitem{fermi1}
  J. Han, L. Zhang, and W. E,
  Solving many-electron Schr\"odinger equation using deep neural networks,
  J. Comput. Phys. \textbf{399}, 108929 (2019).

\bibitem{fermi2}
  K. Choo, A. Mezzacapo, and G. Carleo,
  Fermionic neural-network states for ab-initio electronic structure,
  Nat. Commun. \textbf{11}, 2368 (2020).
  
\bibitem{fermi3}
  D. Pfau, J. S. Spencer, A. G. D. G. Matthews, and W. M. C. Foulkes,
  \textit{Ab initio} solution of the many-electron Schr\"odinger
  equation with deep neural networks,
  Phys. Rev. Res. \textbf{2}, 033429 (2020).

\bibitem{fermi4}
  J. Hermann, A. Sch\"atzle, and F. No\'e,
  Deep-neural-network solution of the electronic Schr\"odinger
  equation,
  Nat. Chem. \textbf{12}, 891 (2020).

\bibitem{fermi5}
  J. W. T. Keeble, M. Drissi, A. Rojo-Franc\'as, B. Juli\'a-D\'iaz,
  and A. Rios,
  Machine learning one-dimensional spinless trapped fermionic systems
  with neural-network quantum states,
  Phys. Rev. A \textbf{108}, 063320 (2023).
  
\bibitem{fermi6}
  M. Wilson, S. Moroni, M. Holzmann, N. Gao, F. Wudarski, T. Vegge,
  and A. Bhowmik,
  Neural network ansatz for periodic wave functions and the
  homogeneous electron gas,
  Phys. Rev. B \textbf{107}, 235139 (2023).

\bibitem{fermi7}
  E. M. Nordhagen, J. M. Kim, B. Fore, A. Lovato, and
  M. Hjorth-Jensen,
  Efficient solutions of fermionic systems using artificial neural
  networks,
  Frontiers Phys. \textbf{11}, 1061580 (2023).

\bibitem{fermi8}
  J. Kim, G. Pescia, B. Fore, J. Nys, G. Carleo, S. Gandolfi,
  M. Hjorth-Jensen, and A. Lovato,
  Neural-network quantum states for ultra-cold Fermi gases,
  Commun. Phys. \textbf{7}, 148 (2024).

\bibitem{nuclei1}
  C. Adams, G. Carleo, A. Lovato, and N. Rocco,
  Variational Monte Carlo Calculations of $A\le 4$ Nuclei with an
  Artificial Neural-Network Correlator Ansatz,
  Phys. Rev. Lett. \textbf{127}, 022502 (2021).

\bibitem{nuclei2}
  A. Gnech, C. Adams, N. Brawand, G. Carleo, A. Lovato, and N. Rocco,
  Nuclei with up to $A=6$ nucleons with artificial neural network wave
  functions,
  Few-Body Syst. \textbf{63}, 7 (2022).

\bibitem{nuclei3}
  A. Lovato, C. Adams, G. Carleo, and N. Rocco,
  Hidden-nucleons neural-network quantum states for the nuclear
  many-body problem,
  Phys. Rev. Res. \textbf{4}, 043178 (2022).

\bibitem{nuclei4}
  Y. L. Yang and P. W. Zhao,
  A consistent description of the relativistic effects and three-body
  interactions in atomic nuclei,
  Phys. Lett. B \textbf{835}, 137587 (2022).

\bibitem{nuclei5}
  Y. L. Yang and P. W. Zhao,
  Deep-neural-network approach to solving the ab initio nuclear
  structure problem,
  Phys. Rev. C \textbf{107}, 034320 (2023).

\bibitem{nuclei6}
  E. Parnes, N. Barnea, G. Carleo, A. Lovato, N. Rocco, and X. Zhang,
  Nuclear Responses with Neural-Network Quantum States,
  Phys. Rev. Lett. \textbf{136}, 032501 (2026).

\bibitem{nuclei7}
  A. D. Donna, L. Contessi, A. Lovato, and F. Pederiva,
  Hypernuclei with neural network quantum states,
  Phys. Rev. Res. \textbf{8}, 013160 (2026).

\bibitem{Choo}
  K. Choo, G. Carleo, N. Regnault, and T. Neupert,
  Symmetries and Many-Body Excitations with Neural-Network Quantum States,
  Phys. Rev. Lett. \textbf{121}, 167204 (2018).

\bibitem{Nomura}
  Y. Nomura,
  Machine Learning Quantum States -- Extensions to Fermion-Boson
  Coupled Systems and Excited-State Calculations,
  J. Phys. Soc. Jpn. \textbf{89}, 054706 (2020).

\bibitem{Viteritti}
  L. L. Viteritti, F. Ferrari, and F. Becca,
  Accuracy of restricted Boltzmann machines for the one-dimensional $J_1$--$J_2$ Heisenberg model,
  SciPost Phys. \textbf{12}, 166 (2022).

\bibitem{Entwistle}
  M. T. Entwistle, Z. Sch\"atzle, P. A. Erdman, J. Hermann, and
  F. No\'e,
  Electronic excited states in deep variational Monte Carlo,
  Nat. Commun. \textbf{14}, 274 (2023).

\bibitem{Pfau}
  D. Pfau, S. Axelrod, H. Sutterud, I. von Glehn, and J. S. Spencer,
  Accurate computation of quantum excited states with neural networks,
  Science \textbf{385}, 846 (2024).

\bibitem{Li}
  Z. Li, Z. Lu, R. Li, X. Wen, X. Li, L. Wang, J. Chen, and W. Ren,
  Spin-symmetry-enforced solution of the many-body Schr\"odinger
  equation with a deep neural network,
  Nat. Comput. Sci. \textbf{4}, 910 (2024).

\bibitem{ring1980nuclear}
  P. Ring and P. Schuck,
  {\it The Nuclear Many-Body Problem},
  (Springer-Verlag, New York, 1980).

\bibitem [{\citenamefont {Inouye}\ \emph {et~al.}(1998)\citenamefont {Inouye},
  \citenamefont {Andrews}, \citenamefont {Stenger}, \citenamefont {Miesner},
  \citenamefont {Stamper-Kurn},\ and\ \citenamefont
  {Ketterle}}]{inouye1998observation}%
  \BibitemOpen
  \bibfield  {author} {\bibinfo {author} {\bibfnamefont {S.}~\bibnamefont
  {Inouye}}, \bibinfo {author} {\bibfnamefont {M.~R.}\ \bibnamefont {Andrews}},
  \bibinfo {author} {\bibfnamefont {J.}~\bibnamefont {Stenger}}, \bibinfo
  {author} {\bibfnamefont {H.~J.}\ \bibnamefont {Miesner}}, \bibinfo {author}
  {\bibfnamefont {D.~M.}\ \bibnamefont {Stamper-Kurn}},\ and\ \bibinfo {author}
  {\bibfnamefont {W.}~\bibnamefont {Ketterle}},\ }\bibfield  {title} {\bibinfo
  {title} {{Observation of {F}eshbach resonances in a {B}ose--{E}instein
  condensate}},\ }\href@noop {} {\bibfield  {journal} {\bibinfo  {journal}
  {Nature (London)}\ }\textbf {\bibinfo {volume} {392}},\ \bibinfo {pages} {151}
  (\bibinfo {year} {1998})}\BibitemShut {NoStop}%
\bibitem [{\citenamefont {Chin}\ \emph {et~al.}(2010)\citenamefont {Chin},
  \citenamefont {Grimm}, \citenamefont {Julienne},\ and\ \citenamefont
  {Tiesinga}}]{chin2010feshbach}%
  \BibitemOpen
  \bibfield  {author} {\bibinfo {author} {\bibfnamefont {C.}~\bibnamefont
  {Chin}}, \bibinfo {author} {\bibfnamefont {R.}~\bibnamefont {Grimm}},
  \bibinfo {author} {\bibfnamefont {P.}~\bibnamefont {Julienne}},\ and\
  \bibinfo {author} {\bibfnamefont {E.}~\bibnamefont {Tiesinga}},\ }\bibfield
  {title} {\bibinfo {title} {Feshbach resonances in ultracold gases},\
  }\href@noop {} {\bibfield  {journal} {\bibinfo  {journal} {Rev. Mod. Phys.}\
  }\textbf {\bibinfo {volume} {82}},\ \bibinfo {pages} {1225} (\bibinfo {year}
  {2010})}\BibitemShut {NoStop}%
\bibitem [{\citenamefont {Efimov}(1970)}]{efimov}%
  \BibitemOpen
  \bibfield  {author} {\bibinfo {author} {\bibfnamefont {V.}~\bibnamefont
  {Efimov}},\ }\bibfield  {title} {\bibinfo {title} {Energy levels arising from
  resonant two-body forces in a three-body system},\ }\href
  {https://doi.org/10.1016/0370-2693(70)90349-7} {\bibfield  {journal}
  {\bibinfo  {journal} {Phys. Lett. B}\ }\textbf {\bibinfo {volume} {33}},\
  \bibinfo {pages} {563} (\bibinfo {year} {1970})}\BibitemShut {NoStop}%
\bibitem [{\citenamefont {Efimov}(1973)}]{efimov1973energy}%
  \BibitemOpen
  \bibfield  {author} {\bibinfo {author} {\bibfnamefont {V.}~\bibnamefont
  {Efimov}},\ }\bibfield  {title} {\bibinfo {title} {{Energy levels of three
  resonantly interacting particles}},\ }\href@noop {} {\bibfield  {journal}
  {\bibinfo  {journal} {Nucl. Phys. A}\ }\textbf {\bibinfo {volume} {210}},\
  \bibinfo {pages} {157} (\bibinfo {year} {1973})}\BibitemShut {NoStop}%
\bibitem [{\citenamefont {Kraemer}\ \emph {et~al.}(2006)\citenamefont
  {Kraemer}, \citenamefont {Mark}, \citenamefont {Waldburger}, \citenamefont
  {Danzl}, \citenamefont {Chin}, \citenamefont {Engeser}, \citenamefont
  {Lange}, \citenamefont {Pilch}, \citenamefont {Jaakkola}, \citenamefont
  {N\"{a}gerl},\ and\ \citenamefont {Grimm}}]{kraemer2006evidence}%
  \BibitemOpen
  \bibfield  {author} {\bibinfo {author} {\bibfnamefont {T.}~\bibnamefont
  {Kraemer}}, \bibinfo {author} {\bibfnamefont {M.}~\bibnamefont {Mark}},
  \bibinfo {author} {\bibfnamefont {P.}~\bibnamefont {Waldburger}}, \bibinfo
  {author} {\bibfnamefont {J.~G.}\ \bibnamefont {Danzl}}, \bibinfo {author}
  {\bibfnamefont {C.}~\bibnamefont {Chin}}, \bibinfo {author} {\bibfnamefont
  {B.}~\bibnamefont {Engeser}}, \bibinfo {author} {\bibfnamefont {A.~D.}\
  \bibnamefont {Lange}}, \bibinfo {author} {\bibfnamefont {K.}~\bibnamefont
  {Pilch}}, \bibinfo {author} {\bibfnamefont {A.}~\bibnamefont {Jaakkola}},
  \bibinfo {author} {\bibfnamefont {H.~C.}\ \bibnamefont {N\"{a}gerl}},\ and\
  \bibinfo {author} {\bibfnamefont {R.}~\bibnamefont {Grimm}},\ }\bibfield
  {title} {\bibinfo {title} {{Evidence for {E}fimov quantum states in an
  ultracold gas of {C}esium atoms}},\ }\href@noop {} {\bibfield  {journal}
  {\bibinfo  {journal} {Nature (London)}\ }\textbf {\bibinfo {volume} {440}},\ \bibinfo
  {pages} {315} (\bibinfo {year} {2006})}\BibitemShut {NoStop}%
\bibitem [{\citenamefont {Braaten}\ and\ \citenamefont
  {Hammer}(2006)}]{Braaten2006259}%
  \BibitemOpen
  \bibfield  {author} {\bibinfo {author} {\bibfnamefont {E.}~\bibnamefont
  {Braaten}}\ and\ \bibinfo {author} {\bibfnamefont {H.-W.}\ \bibnamefont
  {Hammer}},\ }\bibfield  {title} {\bibinfo {title} {Universality in few-body
  systems with large scattering length},\ }\href
  {https://doi.org/https://doi.org/10.1016/j.physrep.2006.03.001} {\bibfield
  {journal} {\bibinfo  {journal} {Phys. Rep.}\ }\textbf {\bibinfo {volume}
  {428}},\ \bibinfo {pages} {259} (\bibinfo {year} {2006})}\BibitemShut
  {NoStop}%
\bibitem [{\citenamefont {Greene}\ \emph {et~al.}(2017)\citenamefont {Greene},
  \citenamefont {Giannakeas},\ and\ \citenamefont
  {P\'erez-R\'{\i}os}}]{RevModPhys.89.035006}%
  \BibitemOpen
  \bibfield  {author} {\bibinfo {author} {\bibfnamefont {C.~H.}\ \bibnamefont
  {Greene}}, \bibinfo {author} {\bibfnamefont {P.}~\bibnamefont {Giannakeas}},\
  and\ \bibinfo {author} {\bibfnamefont {J.}~\bibnamefont
  {P\'erez-R\'{\i}os}},\ }\bibfield  {title} {\bibinfo {title} {Universal
  few-body physics and cluster formation},\ }\href
  {https://doi.org/10.1103/RevModPhys.89.035006} {\bibfield  {journal}
  {\bibinfo  {journal} {Rev. Mod. Phys.}\ }\textbf {\bibinfo {volume} {89}},\
  \bibinfo {pages} {035006} (\bibinfo {year} {2017})}\BibitemShut {NoStop}%
\bibitem [{\citenamefont {Naidon}\ and\ \citenamefont {Endo}(2017)}]{naidon}%
  \BibitemOpen
  \bibfield  {author} {\bibinfo {author} {\bibfnamefont {P.}~\bibnamefont
  {Naidon}}\ and\ \bibinfo {author} {\bibfnamefont {S.}~\bibnamefont {Endo}},\
  }\bibfield  {title} {\bibinfo {title} {Efimov physics: a review},\ }\href
  {https://doi.org/10.1088/1361-6633/aa50e8} {\bibfield  {journal} {\bibinfo
  {journal} {Rep. Prog. Phys.}\ }\textbf {\bibinfo {volume} {80}},\ \bibinfo
  {pages} {056001} (\bibinfo {year} {2017})}\BibitemShut {NoStop}%
\bibitem{d2018few}
  J. P. D'Incao,
  Few-body physics in resonantly interacting ultracold quantum gases,
  J. Phys. B: At. Mol. Opt. Phys. \textbf{51}, 043001 (2018).

\bibitem [{\citenamefont {Endo}\ \emph {et~al.}(2025)\citenamefont {Endo},
  \citenamefont {Epelbaum}, \citenamefont {Naidon}, \citenamefont {Nishida},
  \citenamefont {Sekiguchi},\ and\ \citenamefont
  {Takahashi}}]{endoepelbaumQcl2024}%
  \BibitemOpen
  \bibfield  {author} {\bibinfo {author} {\bibfnamefont {S.}~\bibnamefont
  {Endo}}, \bibinfo {author} {\bibfnamefont {E.}~\bibnamefont {Epelbaum}},
  \bibinfo {author} {\bibfnamefont {P.}~\bibnamefont {Naidon}}, \bibinfo
  {author} {\bibfnamefont {Y.}~\bibnamefont {Nishida}}, \bibinfo {author}
  {\bibfnamefont {K.}~\bibnamefont {Sekiguchi}},\ and\ \bibinfo {author}
  {\bibfnamefont {Y.}~\bibnamefont {Takahashi}},\ }\bibfield  {title} {\bibinfo
  {title} {Three-body forces and {E}fimov physics in nuclei and atoms},\ }\href
  {https://doi.org/https://doi.org/10.1140/epja/s10050-024-01467-4} {\bibfield
  {journal} {\bibinfo  {journal} {Eur. Phys. J. A}\ }\textbf {\bibinfo {volume}
  {61}},\ \bibinfo {pages} {9} (\bibinfo {year} {2025})}\BibitemShut {NoStop}%
\bibitem [{\citenamefont {Bedaque}\ \emph {et~al.}(1999)\citenamefont
  {Bedaque}, \citenamefont {Hammer},\ and\ \citenamefont {van
  Kolck}}]{PhysRevLett.82.463}%
  \BibitemOpen
  \bibfield  {author} {\bibinfo {author} {\bibfnamefont {P.~F.}\ \bibnamefont
  {Bedaque}}, \bibinfo {author} {\bibfnamefont {H.-W.}\ \bibnamefont
  {Hammer}},\ and\ \bibinfo {author} {\bibfnamefont {U.}~\bibnamefont {van
  Kolck}},\ }\bibfield  {title} {\bibinfo {title} {Renormalization of the
  three-body system with short-range interactions},\ }\href
  {https://doi.org/10.1103/PhysRevLett.82.463} {\bibfield  {journal} {\bibinfo
  {journal} {Phys. Rev. Lett.}\ }\textbf {\bibinfo {volume} {82}},\ \bibinfo
  {pages} {463} (\bibinfo {year} {1999})}\BibitemShut {NoStop}%
\bibitem [{\citenamefont {Zaccanti}\ \emph {et~al.}(2009)\citenamefont
  {Zaccanti}, \citenamefont {Deissler}, \citenamefont {D'Errico}, \citenamefont
  {Fattori}, \citenamefont {Jona-Lasinio}, \citenamefont {M{\"u}ller},
  \citenamefont {Roati}, \citenamefont {Inguscio},\ and\ \citenamefont
  {Modugno}}]{Zaccanti2009}%
  \BibitemOpen
  \bibfield  {author} {\bibinfo {author} {\bibfnamefont {M.}~\bibnamefont
  {Zaccanti}}, \bibinfo {author} {\bibfnamefont {B.}~\bibnamefont {Deissler}},
  \bibinfo {author} {\bibfnamefont {C.}~\bibnamefont {D'Errico}}, \bibinfo
  {author} {\bibfnamefont {M.}~\bibnamefont {Fattori}}, \bibinfo {author}
  {\bibfnamefont {M.}~\bibnamefont {Jona-Lasinio}}, \bibinfo {author}
  {\bibfnamefont {S.}~\bibnamefont {M{\"u}ller}}, \bibinfo {author}
  {\bibfnamefont {G.}~\bibnamefont {Roati}}, \bibinfo {author} {\bibfnamefont
  {M.}~\bibnamefont {Inguscio}},\ and\ \bibinfo {author} {\bibfnamefont
  {G.}~\bibnamefont {Modugno}},\ }\bibfield  {title} {\bibinfo {title}
  {Observation of an {E}fimov spectrum in an atomic system},\ }\href
  {https://doi.org/10.1038/nphys1334} {\bibfield  {journal} {\bibinfo
  {journal} {Nat. Phys.}\ }\textbf {\bibinfo {volume} {5}},\ \bibinfo
  {pages} {586} (\bibinfo {year} {2009})}\BibitemShut {NoStop}%
\bibitem [{\citenamefont {Gross}\ \emph {et~al.}(2009)\citenamefont {Gross},
  \citenamefont {Shotan}, \citenamefont {Kokkelmans},\ and\ \citenamefont
  {Khaykovich}}]{PhysRevLett.103.163202}%
  \BibitemOpen
  \bibfield  {author} {\bibinfo {author} {\bibfnamefont {N.}~\bibnamefont
  {Gross}}, \bibinfo {author} {\bibfnamefont {Z.}~\bibnamefont {Shotan}},
  \bibinfo {author} {\bibfnamefont {S.}~\bibnamefont {Kokkelmans}},\ and\
  \bibinfo {author} {\bibfnamefont {L.}~\bibnamefont {Khaykovich}},\ }\bibfield
   {title} {\bibinfo {title} {Observation of universality in ultracold
  $^{7}\mathrm{Li}$ three-body recombination},\ }\href
  {https://doi.org/10.1103/PhysRevLett.103.163202} {\bibfield  {journal}
  {\bibinfo  {journal} {Phys. Rev. Lett.}\ }\textbf {\bibinfo {volume} {103}},\
  \bibinfo {pages} {163202} (\bibinfo {year} {2009})}\BibitemShut {NoStop}%
\bibitem [{\citenamefont {Pollack}\ \emph {et~al.}(2009)\citenamefont
  {Pollack}, \citenamefont {Dries},\ and\ \citenamefont
  {Hulet}}]{doi:10.1126/science.1182840}%
  \BibitemOpen
  \bibfield  {author} {\bibinfo {author} {\bibfnamefont {S.~E.}\ \bibnamefont
  {Pollack}}, \bibinfo {author} {\bibfnamefont {D.}~\bibnamefont {Dries}},\
  and\ \bibinfo {author} {\bibfnamefont {R.~G.}\ \bibnamefont {Hulet}},\
  }\bibfield  {title} {\bibinfo {title} {Universality in three- and four-body
  bound states of ultracold atoms},\ }\href
  {https://doi.org/10.1126/science.1182840} {\bibfield  {journal} {\bibinfo
  {journal} {Science}\ }\textbf {\bibinfo {volume} {326}},\ \bibinfo {pages}
  {1683} (\bibinfo {year} {2009})}\BibitemShut {NoStop}%
\bibitem [{\citenamefont {Ferlaino}\ \emph {et~al.}(2009)\citenamefont
  {Ferlaino}, \citenamefont {Knoop}, \citenamefont {Berninger}, \citenamefont
  {Harm}, \citenamefont {D'Incao}, \citenamefont {N\"agerl},\ and\
  \citenamefont {Grimm}}]{PhysRevLett.102.140401}%
  \BibitemOpen
  \bibfield  {author} {\bibinfo {author} {\bibfnamefont {F.}~\bibnamefont
  {Ferlaino}}, \bibinfo {author} {\bibfnamefont {S.}~\bibnamefont {Knoop}},
  \bibinfo {author} {\bibfnamefont {M.}~\bibnamefont {Berninger}}, \bibinfo
  {author} {\bibfnamefont {W.}~\bibnamefont {Harm}}, \bibinfo {author}
  {\bibfnamefont {J.~P.}\ \bibnamefont {D'Incao}}, \bibinfo {author}
  {\bibfnamefont {H.-C.}\ \bibnamefont {N\"agerl}},\ and\ \bibinfo {author}
  {\bibfnamefont {R.}~\bibnamefont {Grimm}},\ }\bibfield  {title} {\bibinfo
  {title} {Evidence for universal four-body states tied to an {E}fimov
  trimer},\ }\href {https://doi.org/10.1103/PhysRevLett.102.140401} {\bibfield
  {journal} {\bibinfo  {journal} {Phys. Rev. Lett.}\ }\textbf {\bibinfo
  {volume} {102}},\ \bibinfo {pages} {140401} (\bibinfo {year}
  {2009})}\BibitemShut {NoStop}%
\bibitem [{\citenamefont {Zenesini}\ \emph {et~al.}(2013)\citenamefont
  {Zenesini}, \citenamefont {Huang}, \citenamefont {Berninger}, \citenamefont
  {Besler}, \citenamefont {Nﾃ､gerl}, \citenamefont {Ferlaino}, \citenamefont
  {Grimm}, \citenamefont {Greene},\ and\ \citenamefont
  {Stecher}}]{Zenesini_2013}%
  \BibitemOpen
  \bibfield  {author} {\bibinfo {author} {\bibfnamefont {A.}~\bibnamefont
  {Zenesini}}, \bibinfo {author} {\bibfnamefont {B.}~\bibnamefont {Huang}},
  \bibinfo {author} {\bibfnamefont {M.}~\bibnamefont {Berninger}}, \bibinfo
  {author} {\bibfnamefont {S.}~\bibnamefont {Besler}}, \bibinfo {author}
  {\bibfnamefont {H.-C.}\ \bibnamefont {N\"agerl}}, \bibinfo {author}
  {\bibfnamefont {F.}~\bibnamefont {Ferlaino}}, \bibinfo {author}
  {\bibfnamefont {R.}~\bibnamefont {Grimm}}, \bibinfo {author} {\bibfnamefont
  {C.~H.}\ \bibnamefont {Greene}},\ and\ \bibinfo {author} {\bibfnamefont
  {J.~v.}\ \bibnamefont {Stecher}},\ }\bibfield  {title} {\bibinfo {title}
  {Resonant five-body recombination in an ultracold gas of bosonic atoms},\
  }\href {https://doi.org/10.1088/1367-2630/15/4/043040} {\bibfield  {journal}
  {\bibinfo  {journal} {New J. Phys.}\ }\textbf {\bibinfo {volume} {15}},\
  \bibinfo {pages} {043040} (\bibinfo {year} {2013})}\BibitemShut {NoStop}%
\bibitem [{\citenamefont {Pires}\ \emph {et~al.}(2014)\citenamefont {Pires},
  \citenamefont {Ulmanis}, \citenamefont {H\"afner}, \citenamefont {Repp},
  \citenamefont {Arias}, \citenamefont {Kuhnle},\ and\ \citenamefont
  {Weidem\"uller}}]{PhysRevLett.112.250404}%
  \BibitemOpen
  \bibfield  {author} {\bibinfo {author} {\bibfnamefont {R.}~\bibnamefont
  {Pires}}, \bibinfo {author} {\bibfnamefont {J.}~\bibnamefont {Ulmanis}},
  \bibinfo {author} {\bibfnamefont {S.}~\bibnamefont {H\"afner}}, \bibinfo
  {author} {\bibfnamefont {M.}~\bibnamefont {Repp}}, \bibinfo {author}
  {\bibfnamefont {A.}~\bibnamefont {Arias}}, \bibinfo {author} {\bibfnamefont
  {E.~D.}\ \bibnamefont {Kuhnle}},\ and\ \bibinfo {author} {\bibfnamefont
  {M.}~\bibnamefont {Weidem\"uller}},\ }\bibfield  {title} {\bibinfo {title}
  {Observation of {E}fimov resonances in a mixture with extreme mass
  imbalance},\ }\href {https://doi.org/10.1103/PhysRevLett.112.250404}
  {\bibfield  {journal} {\bibinfo  {journal} {Phys. Rev. Lett.}\ }\textbf
  {\bibinfo {volume} {112}},\ \bibinfo {pages} {250404} (\bibinfo {year}
  {2014})}\BibitemShut {NoStop}%
\bibitem [{\citenamefont {Tung}\ \emph {et~al.}(2014)\citenamefont {Tung},
  \citenamefont {Jim\'enez-Garc\'{\i}a}, \citenamefont {Johansen},
  \citenamefont {Parker},\ and\ \citenamefont {Chin}}]{PhysRevLett.113.240402}%
  \BibitemOpen
  \bibfield  {author} {\bibinfo {author} {\bibfnamefont {S.-K.}\ \bibnamefont
  {Tung}}, \bibinfo {author} {\bibfnamefont {K.}~\bibnamefont
  {Jim\'enez-Garc\'{\i}a}}, \bibinfo {author} {\bibfnamefont {J.}~\bibnamefont
  {Johansen}}, \bibinfo {author} {\bibfnamefont {C.~V.}\ \bibnamefont
  {Parker}},\ and\ \bibinfo {author} {\bibfnamefont {C.}~\bibnamefont {Chin}},\
  }\bibfield  {title} {\bibinfo {title} {Geometric scaling of {E}fimov states
  in a $^{6}\mathrm{Li}\text{\ensuremath{-}}^{133}\mathrm{Cs}$ mixture},\
  }\href {https://doi.org/10.1103/PhysRevLett.113.240402} {\bibfield  {journal}
  {\bibinfo  {journal} {Phys. Rev. Lett.}\ }\textbf {\bibinfo {volume} {113}},\
  \bibinfo {pages} {240402} (\bibinfo {year} {2014})}\BibitemShut {NoStop}%
\bibitem [{\citenamefont {Esry}\ \emph {et~al.}(1999)\citenamefont {Esry},
  \citenamefont {Greene},\ and\ \citenamefont {Burke}}]{PhysRevLett.83.1751}%
  \BibitemOpen
  \bibfield  {author} {\bibinfo {author} {\bibfnamefont {B.~D.}\ \bibnamefont
  {Esry}}, \bibinfo {author} {\bibfnamefont {C.~H.}\ \bibnamefont {Greene}},\
  and\ \bibinfo {author} {\bibfnamefont {J.~P.}\ \bibnamefont {Burke}},\
  }\bibfield  {title} {\bibinfo {title} {Recombination of three atoms in the
  ultracold limit},\ }\href {https://doi.org/10.1103/PhysRevLett.83.1751}
  {\bibfield  {journal} {\bibinfo  {journal} {Phys. Rev. Lett.}\ }\textbf
  {\bibinfo {volume} {83}},\ \bibinfo {pages} {1751} (\bibinfo {year}
  {1999})}\BibitemShut {NoStop}%
\bibitem [{\citenamefont {von Stecher}\ \emph {et~al.}(2009)\citenamefont {von
  Stecher}, \citenamefont {D窶僮ncao},\ and\ \citenamefont
  {Greene}}]{von2009signatures}%
  \BibitemOpen
  \bibfield  {author} {\bibinfo {author} {\bibfnamefont {J.}~\bibnamefont {von
  Stecher}}, \bibinfo {author} {\bibfnamefont {J.~P.}\ \bibnamefont
  {D'Incao}},\ and\ \bibinfo {author} {\bibfnamefont {C.~H.}\ \bibnamefont
  {Greene}},\ }\bibfield  {title} {\bibinfo {title} {Signatures of universal
  four-body phenomena and their relation to the {E}fimov effect},\ }\href
  {https://doi.org/https://doi.org/10.1038/nphys1253} {\bibfield  {journal}
  {\bibinfo  {journal} {Nat. Phys.}\ }\textbf {\bibinfo {volume} {5}},\
  \bibinfo {pages} {417} (\bibinfo {year} {2009})}\BibitemShut {NoStop}%
\bibitem [{\citenamefont {von Stecher}(2010)}]{vonStecher2010}%
  \BibitemOpen
  \bibfield  {author} {\bibinfo {author} {\bibfnamefont {J.}~\bibnamefont {von
  Stecher}},\ }\bibfield  {title} {\bibinfo {title} {Weakly bound cluster
  states of {E}fimov character},\ }\href
  {https://doi.org/10.1088/0953-4075/43/10/101002} {\bibfield  {journal}
  {\bibinfo  {journal} {J. Phys. B}\ }\textbf {\bibinfo {volume} {43}},\
  \bibinfo {pages} {101002} (\bibinfo {year} {2010})}\BibitemShut {NoStop}%
\bibitem [{\citenamefont {Gattobigio}\ and\ \citenamefont
  {Kievsky}(2014)}]{finite}%
  \BibitemOpen
  \bibfield  {author} {\bibinfo {author} {\bibfnamefont {M.}~\bibnamefont
  {Gattobigio}}\ and\ \bibinfo {author} {\bibfnamefont {A.}~\bibnamefont
  {Kievsky}},\ }\bibfield  {title} {\bibinfo {title} {Universality and scaling
  in the $n$-body sector of {E}fimov physics},\ }\href
  {https://doi.org/10.1103/PhysRevA.90.012502} {\bibfield  {journal} {\bibinfo
  {journal} {Phys. Rev. A}\ }\textbf {\bibinfo {volume} {90}},\ \bibinfo
  {pages} {012502} (\bibinfo {year} {2014})}\BibitemShut {NoStop}%
\bibitem [{\citenamefont {Naidon}\ \emph {et~al.}(2012)\citenamefont {Naidon},
  \citenamefont {Hiyama},\ and\ \citenamefont {Ueda}}]{PhysRevA.86.012502}%
  \BibitemOpen
  \bibfield  {author} {\bibinfo {author} {\bibfnamefont {P.}~\bibnamefont
  {Naidon}}, \bibinfo {author} {\bibfnamefont {E.}~\bibnamefont {Hiyama}},\
  and\ \bibinfo {author} {\bibfnamefont {M.}~\bibnamefont {Ueda}},\ }\bibfield
  {title} {\bibinfo {title} {Universality and the three-body parameter of
  ${}^{4}${H}e trimers},\ }\href {https://doi.org/10.1103/PhysRevA.86.012502}
  {\bibfield  {journal} {\bibinfo  {journal} {Phys. Rev. A}\ }\textbf {\bibinfo
  {volume} {86}},\ \bibinfo {pages} {012502} (\bibinfo {year}
  {2012})}\BibitemShut {NoStop}%
\bibitem [{\citenamefont {Hiyama}\ and\ \citenamefont
  {Kamimura}(2014)}]{PhysRevA.90.052514}%
  \BibitemOpen
  \bibfield  {author} {\bibinfo {author} {\bibfnamefont {E.}~\bibnamefont
  {Hiyama}}\ and\ \bibinfo {author} {\bibfnamefont {M.}~\bibnamefont
  {Kamimura}},\ }\bibfield  {title} {\bibinfo {title} {Universality in
  {E}fimov-associated tetramers in $^{4}\mathrm{He}$},\ }\href
  {https://doi.org/10.1103/PhysRevA.90.052514} {\bibfield  {journal} {\bibinfo
  {journal} {Phys. Rev. A}\ }\textbf {\bibinfo {volume} {90}},\ \bibinfo
  {pages} {052514} (\bibinfo {year} {2014})}\BibitemShut {NoStop}%
\bibitem [{\citenamefont {Grebenev}\ \emph {et~al.}(1998)\citenamefont
  {Grebenev}, \citenamefont {Toennies},\ and\ \citenamefont
  {Vilesov}}]{science.279.5359.2083}%
  \BibitemOpen
  \bibfield  {author} {\bibinfo {author} {\bibfnamefont {S.}~\bibnamefont
  {Grebenev}}, \bibinfo {author} {\bibfnamefont {J.~P.}\ \bibnamefont
  {Toennies}},\ and\ \bibinfo {author} {\bibfnamefont {A.~F.}\ \bibnamefont
  {Vilesov}},\ }\bibfield  {title} {\bibinfo {title} {Superfluidity within a
  small helium-4 cluster: The microscopic andronikashvili experiment},\ }\href
  {https://doi.org/10.1126/science.279.5359.2083} {\bibfield  {journal}
  {\bibinfo  {journal} {Science}\ }\textbf {\bibinfo {volume} {279}},\ \bibinfo
  {pages} {2083} (\bibinfo {year} {1998})}\BibitemShut {NoStop}%
\bibitem{Wenz_2013}
  A. N. Wenz, G. Z\"urn, S. Murmann, I. Brouzos, T. Lompe, and
  S. Jochim,
  From Few to Many: Observing the Formation of a Fermi Sea One Atom at
  a Time,
  Science \textbf{342}, 457 (2013).

\bibitem [{\citenamefont {Yan}\ and\ \citenamefont
  {Blume}(2015)}]{PhysRevA.92.033626}%
  \BibitemOpen
  \bibfield  {author} {\bibinfo {author} {\bibfnamefont {Y.}~\bibnamefont
  {Yan}}\ and\ \bibinfo {author} {\bibfnamefont {D.}~\bibnamefont {Blume}},\
  }\bibfield  {title} {\bibinfo {title} {Energy and structural properties of
  $n$-boson clusters attached to three-body {E}fimov states: Two-body
  zero-range interactions and the role of the three-body regulator},\ }\href
  {https://doi.org/10.1103/PhysRevA.92.033626} {\bibfield  {journal} {\bibinfo
  {journal} {Phys. Rev. A}\ }\textbf {\bibinfo {volume} {92}},\ \bibinfo
  {pages} {033626} (\bibinfo {year} {2015})}\BibitemShut {NoStop}%
\bibitem [{\citenamefont {Levinsen}\ \emph {et~al.}(2017)\citenamefont
  {Levinsen}, \citenamefont {Massignan}, \citenamefont {Endo},\ and\
  \citenamefont {Parish}}]{Levinsen_2017}%
  \BibitemOpen
  \bibfield  {author} {\bibinfo {author} {\bibfnamefont {J.}~\bibnamefont
  {Levinsen}}, \bibinfo {author} {\bibfnamefont {P.}~\bibnamefont {Massignan}},
  \bibinfo {author} {\bibfnamefont {S.}~\bibnamefont {Endo}},\ and\ \bibinfo
  {author} {\bibfnamefont {M.~M.}\ \bibnamefont {Parish}},\ }\bibfield  {title}
  {\bibinfo {title} {Universality of the unitary {F}ermi gas: a few-body
  perspective},\ }\href {https://doi.org/10.1088/1361-6455/aa5a1e} {\bibfield
  {journal} {\bibinfo  {journal} {J. Phys. B}\ }\textbf {\bibinfo {volume}
  {50}},\ \bibinfo {pages} {072001} (\bibinfo {year} {2017})}\BibitemShut
  {NoStop}%
\bibitem [{\citenamefont {Blume}\ and\ \citenamefont {Daily}(2010)}]{Blume}%
  \BibitemOpen
  \bibfield  {author} {\bibinfo {author} {\bibfnamefont {D.}~\bibnamefont
  {Blume}}\ and\ \bibinfo {author} {\bibfnamefont {K.~M.}\ \bibnamefont
  {Daily}},\ }\bibfield  {title} {\bibinfo {title} {Breakdown of universality
  for unequal-mass {F}ermi gases with infinite scattering length},\ }\href
  {https://doi.org/10.1103/PhysRevLett.105.170403} {\bibfield  {journal}
  {\bibinfo  {journal} {Phys. Rev. Lett.}\ }\textbf {\bibinfo {volume} {105}},\
  \bibinfo {pages} {170403} (\bibinfo {year} {2010})}\BibitemShut {NoStop}%
\bibitem [{\citenamefont {Daily}\ and\ \citenamefont {Blume}(2012)}]{Daily}%
  \BibitemOpen
  \bibfield  {author} {\bibinfo {author} {\bibfnamefont {K.~M.}\ \bibnamefont
  {Daily}}\ and\ \bibinfo {author} {\bibfnamefont {D.}~\bibnamefont {Blume}},\
  }\bibfield  {title} {\bibinfo {title} {Thermodynamics of the two-component
  {F}ermi gas with unequal masses at unitarity},\ }\href
  {https://doi.org/10.1103/PhysRevA.85.013609} {\bibfield  {journal} {\bibinfo
  {journal} {Phys. Rev. A}\ }\textbf {\bibinfo {volume} {85}},\ \bibinfo
  {pages} {013609} (\bibinfo {year} {2012})}\BibitemShut {NoStop}%
\bibitem [{\citenamefont {Hammer}\ and\ \citenamefont
  {Platter}(2010)}]{AnnRev_HamPlatt}%
  \BibitemOpen
  \bibfield  {author} {\bibinfo {author} {\bibfnamefont {H.~W.}\ \bibnamefont
  {Hammer}}\ and\ \bibinfo {author} {\bibfnamefont {L.}~\bibnamefont
  {Platter}},\ }\bibfield  {title} {\bibinfo {title} {Efimov states in nuclear
  and particle physics},\ }\href@noop {} {\bibfield  {journal} {\bibinfo
  {journal} {Ann. Rev. Nucl. Part. Sci.}\ }\textbf {\bibinfo {volume} {60}},\
  \bibinfo {pages} {207} (\bibinfo {year} {2010})}\BibitemShut {NoStop}%
\bibitem [{\citenamefont {Tanihata}\ \emph {et~al.}(2008)\citenamefont
  {Tanihata}, \citenamefont {Alcorta}, \citenamefont {Bandyopadhyay},
  \citenamefont {Bieri}, \citenamefont {Buchmann}, \citenamefont {Davids},
  \citenamefont {Galinski}, \citenamefont {Howell}, \citenamefont {Mills},
  \citenamefont {Mythili}, \citenamefont {Openshaw}, \citenamefont
  {Padilla-Rodal}, \citenamefont {Ruprecht}, \citenamefont {Sheffer},
  \citenamefont {Shotter}, \citenamefont {Trinczek}, \citenamefont {Walden},
  \citenamefont {Savajols}, \citenamefont {Roger}, \citenamefont {Caamano},
  \citenamefont {Mittig}, \citenamefont {Roussel-Chomaz}, \citenamefont
  {Kanungo}, \citenamefont {Gallant}, \citenamefont {Notani}, \citenamefont
  {Savard},\ and\ \citenamefont {Thompson}}]{PhysRevLett.100.192502}%
  \BibitemOpen
  \bibfield  {author} {\bibinfo {author} {\bibfnamefont {I.}~\bibnamefont
  {Tanihata}}, \bibinfo {author} {\bibfnamefont {M.}~\bibnamefont {Alcorta}},
  \bibinfo {author} {\bibfnamefont {D.}~\bibnamefont {Bandyopadhyay}}, \bibinfo
  {author} {\bibfnamefont {R.}~\bibnamefont {Bieri}}, \bibinfo {author}
  {\bibfnamefont {L.}~\bibnamefont {Buchmann}}, \bibinfo {author}
  {\bibfnamefont {B.}~\bibnamefont {Davids}}, \bibinfo {author} {\bibfnamefont
  {N.}~\bibnamefont {Galinski}}, \bibinfo {author} {\bibfnamefont
  {D.}~\bibnamefont {Howell}}, \bibinfo {author} {\bibfnamefont
  {W.}~\bibnamefont {Mills}}, \bibinfo {author} {\bibfnamefont
  {S.}~\bibnamefont {Mythili}}, \bibinfo {author} {\bibfnamefont
  {R.}~\bibnamefont {Openshaw}}, \bibinfo {author} {\bibfnamefont
  {E.}~\bibnamefont {Padilla-Rodal}}, \bibinfo {author} {\bibfnamefont
  {G.}~\bibnamefont {Ruprecht}}, \bibinfo {author} {\bibfnamefont
  {G.}~\bibnamefont {Sheffer}}, \bibinfo {author} {\bibfnamefont {A.~C.}\
  \bibnamefont {Shotter}}, \bibinfo {author} {\bibfnamefont {M.}~\bibnamefont
  {Trinczek}}, \bibinfo {author} {\bibfnamefont {P.}~\bibnamefont {Walden}},
  \bibinfo {author} {\bibfnamefont {H.}~\bibnamefont {Savajols}}, \bibinfo
  {author} {\bibfnamefont {T.}~\bibnamefont {Roger}}, \bibinfo {author}
  {\bibfnamefont {M.}~\bibnamefont {Caamano}}, \bibinfo {author} {\bibfnamefont
  {W.}~\bibnamefont {Mittig}}, \bibinfo {author} {\bibfnamefont
  {P.}~\bibnamefont {Roussel-Chomaz}}, \bibinfo {author} {\bibfnamefont
  {R.}~\bibnamefont {Kanungo}}, \bibinfo {author} {\bibfnamefont
  {A.}~\bibnamefont {Gallant}}, \bibinfo {author} {\bibfnamefont
  {M.}~\bibnamefont {Notani}}, \bibinfo {author} {\bibfnamefont
  {G.}~\bibnamefont {Savard}},\ and\ \bibinfo {author} {\bibfnamefont {I.~J.}\
  \bibnamefont {Thompson}},\ }\bibfield  {title} {\bibinfo {title} {Measurement
  of the two-halo neutron transfer reaction
  $^{1}\mathrm{H}(^{11}\mathrm{Li},^{9}\mathrm{Li})^{3}\mathrm{H}$ at $3a\text{
  }\text{ }\mathrm{MeV}$},\ }\href
  {https://doi.org/10.1103/PhysRevLett.100.192502} {\bibfield  {journal}
  {\bibinfo  {journal} {Phys. Rev. Lett.}\ }\textbf {\bibinfo {volume} {100}},\
  \bibinfo {pages} {192502} (\bibinfo {year} {2008})}\BibitemShut {NoStop}%
\bibitem [{\citenamefont {Hove}\ \emph {et~al.}(2018)\citenamefont {Hove},
  \citenamefont {Garrido}, \citenamefont {Sarriguren}, \citenamefont {Fedorov},
  \citenamefont {Fynbo}, \citenamefont {Jensen},\ and\ \citenamefont
  {Zinner}}]{PhysRevLett.120.052502}%
  \BibitemOpen
  \bibfield  {author} {\bibinfo {author} {\bibfnamefont {D.}~\bibnamefont
  {Hove}}, \bibinfo {author} {\bibfnamefont {E.}~\bibnamefont {Garrido}},
  \bibinfo {author} {\bibfnamefont {P.}~\bibnamefont {Sarriguren}}, \bibinfo
  {author} {\bibfnamefont {D.~V.}\ \bibnamefont {Fedorov}}, \bibinfo {author}
  {\bibfnamefont {H.~O.~U.}\ \bibnamefont {Fynbo}}, \bibinfo {author}
  {\bibfnamefont {A.~S.}\ \bibnamefont {Jensen}},\ and\ \bibinfo {author}
  {\bibfnamefont {N.~T.}\ \bibnamefont {Zinner}},\ }\bibfield  {title}
  {\bibinfo {title} {},\ }\href
  {https://doi.org/10.1103/PhysRevLett.120.052502} {\bibfield  {journal}
  {\bibinfo  {journal} {Phys. Rev. Lett.}\ }\textbf {\bibinfo {volume} {120}},\
  \bibinfo {pages} {052502} (\bibinfo {year} {2018})}\BibitemShut {NoStop}%
\bibitem{Hove}
  D. Hove, E. Garrido, P. Sarriguren, D. V. Fedorov, H. O. U. Fynbo,
  A. S. Jensen, and N. T. Zinner,
  Emergence of clusters: Halos, Efimov states, and experimental
  signals,
  Phys. Rev. Lett. \textbf{120}, 052502 (2018).

\bibitem{Kunitski}
  M. Kunitski, S. Zeller, J. Voigtsberger, A. Kalinin,
  L. P. H. Schmidt, M. Sch\"offler, A. Czasch, W. Sch\"ollkopf,
  R. E. Grisenti, T. Jahnke, D. Blume, and R. D\"orner,
  Observation of the Efimov state of the helium trimer,
  Science \textbf{348}, 551 (2015).
  
\bibitem [{\citenamefont {Nishida}\ \emph {et~al.}(2013)\citenamefont
  {Nishida}, \citenamefont {Kato},\ and\ \citenamefont
  {Batista}}]{nishida2013efimov}%
  \BibitemOpen
  \bibfield  {author} {\bibinfo {author} {\bibfnamefont {Y.}~\bibnamefont
  {Nishida}}, \bibinfo {author} {\bibfnamefont {Y.}~\bibnamefont {Kato}},\ and\
  \bibinfo {author} {\bibfnamefont {C.~D.}\ \bibnamefont {Batista}},\
  }\bibfield  {title} {\bibinfo {title} {Efimov effect in quantum magnets},\
  }\href {https://doi.org/10.1038/nphys2523} {\bibfield  {journal} {\bibinfo
  {journal} {Nat. Phys.}\ }\textbf {\bibinfo {volume} {9}},\ \bibinfo
  {pages} {93} (\bibinfo {year} {2013})}\BibitemShut {NoStop}%
\bibitem [{\citenamefont {Petrov}(2003)}]{PhysRevA.67.010703}%
  \BibitemOpen
  \bibfield  {author} {\bibinfo {author} {\bibfnamefont {D.~S.}\ \bibnamefont
  {Petrov}},\ }\bibfield  {title} {\bibinfo {title} {Three-body problem in
  {F}ermi gases with short-range interparticle interaction},\ }\href
  {https://doi.org/10.1103/PhysRevA.67.010703} {\bibfield  {journal} {\bibinfo
  {journal} {Phys. Rev. A}\ }\textbf {\bibinfo {volume} {67}},\ \bibinfo
  {pages} {010703} (\bibinfo {year} {2003})}\BibitemShut {NoStop}%
\bibitem{Goodfellow}
  I. Goodfellow, Y. Bengio, and A. Courville,
  {\it Deep Learning},
  (MIT Press, Cambridge, 2016).
\bibitem{Adam}
  D. P. Kingma and J. Ba,
  Adam: A Method for Stochastic Optimization,
  arXiv:1412.6980.
\bibitem [{\citenamefont {P{\"o}schl}\ and\ \citenamefont
  {Teller}(1933)}]{PoschlTeller1933}%
  \BibitemOpen
  \bibfield  {author} {\bibinfo {author} {\bibfnamefont {G.}~\bibnamefont
  {P{\"o}schl}}\ and\ \bibinfo {author} {\bibfnamefont {E.}~\bibnamefont
  {Teller}},\ }\bibfield  {title} {\bibinfo {title} {Bemerkungen zur
  quantenmechanik des anharmonischen oszillators},\ }\href
  {https://doi.org/10.1007/BF01331132} {\bibfield  {journal} {\bibinfo
  {journal} {Z. Phys.}\ }\textbf {\bibinfo {volume} {83}},\ \bibinfo {pages}
  {143} (\bibinfo {year} {1933})}\BibitemShut {NoStop}%
\bibitem{flugge2012practical}
  S. Fl\"ugge,
  {\it Practical Quantum Mechanics},
  (Springer-Verlag, New York, 1974).
\bibitem [{\citenamefont {Sch{\"a}fer}\ \emph {et~al.}(2023)\citenamefont
  {Sch{\"a}fer}, \citenamefont {Haruna},\ and\ \citenamefont {Takahashi}}]{er}%
  \BibitemOpen
  \bibfield  {author} {\bibinfo {author} {\bibfnamefont {F.}~\bibnamefont
  {Sch{\"a}fer}}, \bibinfo {author} {\bibfnamefont {Y.}~\bibnamefont
  {Haruna}},\ and\ \bibinfo {author} {\bibfnamefont {Y.}~\bibnamefont
  {Takahashi}},\ }\bibfield  {title} {\bibinfo {title} {Observation of
  {F}eshbach resonances in an $^{167}${E}r--$^{6}${L}i {F}ermi--{F}ermi
  mixture},\ }\href {https://doi.org/10.7566/JPSJ.92.054301} {\bibfield
  {journal} {\bibinfo  {journal} {J. Phys. Soc. Jpn.}\ }\textbf {\bibinfo
  {volume} {92}},\ \bibinfo {pages} {054301} (\bibinfo {year}
  {2023})}\BibitemShut {NoStop}%
\bibitem [{\citenamefont {Xie}\ \emph {et~al.}(2025)\citenamefont {Xie},
  \citenamefont {Li}, \citenamefont {Zhou}, \citenamefont {Luo}, \citenamefont
  {Wang}, \citenamefont {Nie}, \citenamefont {Shen}, \citenamefont {Chen},
  \citenamefont {Yao},\ and\ \citenamefont {Pan}}]{DyLifeshbach}%
  \BibitemOpen
  \bibfield  {author} {\bibinfo {author} {\bibfnamefont {K.}~\bibnamefont
  {Xie}}, \bibinfo {author} {\bibfnamefont {X.}~\bibnamefont {Li}}, \bibinfo
  {author} {\bibfnamefont {Y.-Y.}\ \bibnamefont {Zhou}}, \bibinfo {author}
  {\bibfnamefont {J.-H.}\ \bibnamefont {Luo}}, \bibinfo {author} {\bibfnamefont
  {S.}~\bibnamefont {Wang}}, \bibinfo {author} {\bibfnamefont {Y.-Z.}\
  \bibnamefont {Nie}}, \bibinfo {author} {\bibfnamefont {H.-C.}\ \bibnamefont
  {Shen}}, \bibinfo {author} {\bibfnamefont {Y.-A.}\ \bibnamefont {Chen}},
  \bibinfo {author} {\bibfnamefont {X.-C.}\ \bibnamefont {Yao}},\ and\ \bibinfo
  {author} {\bibfnamefont {J.-W.}\ \bibnamefont {Pan}},\ }\bibfield  {title}
  {\bibinfo {title} {Feshbach spectroscopy of ultracold mixtures of
  $^{6}\mathrm{Li}$ and $^{164}\mathrm{Dy}$ atoms},\ }\href
  {https://doi.org/10.1103/PhysRevA.111.023327} {\bibfield  {journal} {\bibinfo
   {journal} {Phys. Rev. A}\ }\textbf {\bibinfo {volume} {111}},\ \bibinfo
  {pages} {023327} (\bibinfo {year} {2025})}\BibitemShut {NoStop}%
\bibitem [{\citenamefont {Oi}\ \emph {et~al.}(2024)\citenamefont {Oi},
  \citenamefont {Naidon},\ and\ \citenamefont {Endo}}]{OiEndo2024}%
  \BibitemOpen
  \bibfield  {author} {\bibinfo {author} {\bibfnamefont {K.}~\bibnamefont
  {Oi}}, \bibinfo {author} {\bibfnamefont {P.}~\bibnamefont {Naidon}},\ and\
  \bibinfo {author} {\bibfnamefont {S.}~\bibnamefont {Endo}},\ }\bibfield
  {title} {\bibinfo {title} {Universality of {E}fimov states in highly
  mass-imbalanced cold-atom mixtures with van der {W}aals and dipole
  interactions},\ }\href {https://doi.org/10.1103/PhysRevA.110.033305}
  {\bibfield  {journal} {\bibinfo  {journal} {Phys. Rev. A}\ }\textbf {\bibinfo
  {volume} {110}},\ \bibinfo {pages} {033305} (\bibinfo {year}
  {2024})}\BibitemShut {NoStop}%
\bibitem [{\citenamefont {Oi}\ and\ \citenamefont
  {Endo}(2025)}]{oi2025universal}%
  \BibitemOpen
  \bibfield  {author} {\bibinfo {author} {\bibfnamefont {K.}~\bibnamefont
  {Oi}}\ and\ \bibinfo {author} {\bibfnamefont {S.}~\bibnamefont {Endo}},\
  }\bibfield  {title} {\bibinfo {title} {Universal {E}fimov spectra and
  fermionic doublets in highly mass-imbalanced cold-atom mixtures with van der
  {W}aals and dipole interactions},\ }\href {https://doi.org/10.1103/z5rx-51yx}
  {\bibfield  {journal} {\bibinfo  {journal} {Phys. Rev. Res.}\ }\textbf
  {\bibinfo {volume} {7}},\ \bibinfo {pages} {033236} (\bibinfo {year}
  {2025})}\BibitemShut {NoStop}%
\bibitem [{\citenamefont {Breit}(1930)}]{breit}%
  \BibitemOpen
  \bibfield  {author} {\bibinfo {author} {\bibfnamefont {G.}~\bibnamefont
  {Breit}},\ }\bibfield  {title} {\bibinfo {title} {Separation of angles in the
  two-electron problem},\ }\href {https://doi.org/10.1103/PhysRev.35.569}
  {\bibfield  {journal} {\bibinfo  {journal} {Phys. Rev.}\ }\textbf {\bibinfo
  {volume} {35}},\ \bibinfo {pages} {569} (\bibinfo {year} {1930})}\BibitemShut
  {NoStop}%
\bibitem [{\citenamefont {Wang}\ \emph
  {et~al.}(2011{\natexlab{a}})\citenamefont {Wang}, \citenamefont {D'Incao},\
  and\ \citenamefont {Greene}}]{PhysRevLett.106.233201}%
  \BibitemOpen
  \bibfield  {author} {\bibinfo {author} {\bibfnamefont {Y.}~\bibnamefont
  {Wang}}, \bibinfo {author} {\bibfnamefont {J.~P.}\ \bibnamefont {D'Incao}},\
  and\ \bibinfo {author} {\bibfnamefont {C.~H.}\ \bibnamefont {Greene}},\
  }\bibfield  {title} {\bibinfo {title} {Efimov effect for three interacting
  bosonic dipoles},\ }\href {https://doi.org/10.1103/PhysRevLett.106.233201}
  {\bibfield  {journal} {\bibinfo  {journal} {Phys. Rev. Lett.}\ }\textbf
  {\bibinfo {volume} {106}},\ \bibinfo {pages} {233201} (\bibinfo {year}
  {2011}{\natexlab{a}})}\BibitemShut {NoStop}%
\bibitem [{\citenamefont {Wang}\ \emph
  {et~al.}(2011{\natexlab{b}})\citenamefont {Wang}, \citenamefont {D'Incao},\
  and\ \citenamefont {Greene}}]{PhysRevLett.107.233201}%
  \BibitemOpen
  \bibfield  {author} {\bibinfo {author} {\bibfnamefont {Y.}~\bibnamefont
  {Wang}}, \bibinfo {author} {\bibfnamefont {J.~P.}\ \bibnamefont {D'Incao}},\
  and\ \bibinfo {author} {\bibfnamefont {C.~H.}\ \bibnamefont {Greene}},\
  }\bibfield  {title} {\bibinfo {title} {Universal three-body physics for
  fermionic dipoles},\ }\href {https://doi.org/10.1103/PhysRevLett.107.233201}
  {\bibfield  {journal} {\bibinfo  {journal} {Phys. Rev. Lett.}\ }\textbf
  {\bibinfo {volume} {107}},\ \bibinfo {pages} {233201} (\bibinfo {year}
  {2011}{\natexlab{b}})}\BibitemShut {NoStop}%
\bibitem [{\citenamefont {Shi}\ \emph {et~al.}(2026)\citenamefont {Shi},
  \citenamefont {Wang},\ and\ \citenamefont {Cui}}]{hcwf-tk6c}%
  \BibitemOpen
  \bibfield  {author} {\bibinfo {author} {\bibfnamefont {T.}~\bibnamefont
  {Shi}}, \bibinfo {author} {\bibfnamefont {H.}~\bibnamefont {Wang}},\ and\
  \bibinfo {author} {\bibfnamefont {X.}~\bibnamefont {Cui}},\ }\bibfield
  {title} {\bibinfo {title} {Universal bound states with {B}ose-{F}ermi duality
  in microwave-shielded ultracold molecules},\ }\href
  {https://doi.org/10.1103/hcwf-tk6c} {\bibfield  {journal} {\bibinfo
  {journal} {Phys. Rev. Lett.}\ }\textbf {\bibinfo {volume} {136}},\ \bibinfo
  {pages} {043402} (\bibinfo {year} {2026})}\BibitemShut {NoStop}%
\bibitem{Ohishi}
  Y. Ohishi, K. Oi, and S. Endo,
  Analytical solution of the Schr\"{o}dinger
  equation with $1/r^3$ and attractive $1/r^2$ potentials: Universal three-body
  parameter of mixed-dimensional Efimov states,
  arXiv:2601.19517.
\end{thebibliography}
\end{document}